\documentclass[12pt,preprint]{aastex}

\newcommand\colheadTwo[1]{\multicolumn{2}{c}{#1}\pt@addcol}%

\slugcomment{Accepted to ApJ on July 7, 2005}

\shorttitle{Chandra and SOFI Observation of Galactic Plane}
\shortauthors{Ebisawa et al.}

\begin{document}


\title{{\it Chandra}\/ Deep X-ray Observation of a Typical Galactic Plane Region and  Near-Infrared
 Identification}


\author{K. Ebisawa\altaffilmark{1,2,3}, M. Tsujimoto\altaffilmark{4}, A. Paizis\altaffilmark{2,5}, 
K. Hamaguchi\altaffilmark{1}, A. Bamba\altaffilmark{6},  R. Cutri\altaffilmark{7}, \\
H. Kaneda\altaffilmark{8},  Y. Maeda\altaffilmark{8}, 
 G. Sato\altaffilmark{8}, A. Senda\altaffilmark{9},
M. Ueno\altaffilmark{9}, S. Yamauchi\altaffilmark{10},  V. Beckmann\altaffilmark{1},\\
T. J.-L. Courvoisier\altaffilmark{2,11}, P. Dubath\altaffilmark{2,11}, \&  E. Nishihara\altaffilmark{12}}
\altaffiltext{1}{code 662, NASA/GSFC, Greenbelt, MD 20771}
\email{ebisawa@milkyway.gsfc.nasa.gov}
\altaffiltext{2}{INTEGRAL Science Data Centre, Chemin d'\'Ecogia 16, 1290, Versoix, Switzerland}
\altaffiltext{3}{Universities Space Research Association}
\altaffiltext{4}{Department of Astronomy and Astrophysics, Pennsylvania State University, University Park,  PA 16802}
\altaffiltext{5}{IASF, Sezione de Milano, via Bassini 15, I-20133, Milano, Italy}
\altaffiltext{6}{Cosmic Radiation Laboratory, RIKEN (The Institute of Physical and Chemical Research), 2--1, Hirosawa, Wako,  Saitama 351-0198, Japan}
\altaffiltext{7}{IPAC, California Institute of Technology, Mail Code 100-22, 770 South Wilson Avenue, Pasadena, CA 91125} 
\altaffiltext{8}{Institute of Space and Astronautical Science/JAXA, Yoshinodai, Sagamihara, Kanagawa, 229-8510 Japan}
\altaffiltext{9}{Department of Physics, Kyoto University, Kitashirakawa Oiwake-cho, Sakyo-ku, Kyoto, 606-8502, Japan}
\altaffiltext{10}{Faculty of Humanities and Social Sciences, Iwate University, Iwate, 020-8550, Japan}
\altaffiltext{11}{Observatory of Geneva, 51 chemin des Maillettes, 1290 Sauverny, Switzerland}
\altaffiltext{12}{Gunma Astronomical Observatory, 6860-86 Nakayama Takayama-mura Agatsuma-gun Gunma, 377-0702, Japan}


\begin{abstract}

Using the {\it Chandra} Advanced CCD Imaging Spectrometer Imaging array (ACIS-I), we have carried out a deep hard X-ray 
observation of the
Galactic plane region at  $(l,b) \approx (28.^\circ5,  0.^\circ0)$, where no discrete X-ray source had been
reported previously.  We have detected 274 new point X-ray sources (4 $\sigma$ confidence) 
as well as strong Galactic diffuse emission
within two partially overlapping ACIS-I  fields ($\sim$ 250 arcmin$^2$ in total).
The point source sensitivity was   $\sim3 \times 10^{-15} $  erg s$^{-1}$ cm$^{-2}$ in the hard X-ray band 
(2 -- 10 keV) and   $\sim2 \times 10^{-16} $ erg s$^{-1}$ cm$^{-2}$ in the soft band (0.5 -- 2 keV).
Sum of all the detected point source fluxes accounts for only $\sim$  10 \% of the total X-ray flux in the field of view.
Even hypothesizing a new population of much dimmer and numerous Galactic point sources,
the total observed X-ray flux cannot be  explained.
Therefore, we conclude that X-ray emission from the Galactic plane has truly diffuse origin.
Removing point sources brighter than  $\sim3 \times 10^{-15} $  erg s$^{-1}$ cm$^{-2}$ (2-- 10 keV),
we have determined the Galactic diffuse X-ray flux as $6.5 \times 10^{-11}$ erg s$^{-1}$ cm$^{-2}$ deg$^{-2}$
(2--10 keV).
Only 26 point sources were detected both in the soft and hard bands, indicating that
there are two distinct classes of the  X-ray sources distinguished by the
spectral hardness ratio.  Surface number density of the hard sources is only slightly
higher than that measured at the high Galactic latitude regions, indicating that
majority of the hard sources are  background AGNs.
Following up the {\it Chandra} observation, we have performed  a near-infrared (NIR) 
survey with SOFI at ESO/NTT.  Almost all the soft X-ray sources have been identified in NIR
and their spectral types are consistent with main-sequence stars, suggesting most of them
are nearby X-ray active stars.
On the other hand, only \mbox{ 22 \%} of the hard sources had NIR counterparts, 
which are presumably Galactic. From X-ray and NIR spectral study, they
are most likely to be quiescent cataclysmic variables.  Our observation suggests 
a population of $\gtrsim 10^4 $ cataclysmic variables  in the entire Galactic plane fainter than $\sim 2 \times
10^{33}$ erg s$^{-1}$.
We have carried out a precise spectral study of the Galactic diffuse X-ray emission 
excluding the point sources.
Confirming previous results, we have detected
prominent emission lines from highly ionized heavy elements in the diffuse
emission.  In particular, central energy of the iron emission  line
was determined  as  $6.52^{+0.08}_{-0.14}$ keV (90 \% confidence), which is significantly lower than what is expected
from a plasma in thermal equilibrium.
The downward shift of the iron line center energy suggests 
non-equilibrium ionization states of the plasma, 
or presence of the non-thermal process to produce  6.4 keV  fluorescent lines.

\end{abstract}




\section{INTRODUCTION}

The Galactic X-ray source population  has been studied from the very beginning
of X-ray astronomy.  The {\it Uhuru}\/ satellite  detected 339 X-ray sources
all over the sky brighter than $\sim2 \times 10^{-12}$ erg s$^{-1}$ cm$^{-2}$
in 2 -- 10 keV (Forman et al.\ 1978).  Most bright X-ray sources are concentrated
near the Galactic bulge and/or distributed along the Galactic plane, indicating  their
Galactic origin.  A high sensitive X-ray observation with
direct imaging was  made for the first time   with the {\it Einstein}\/ satellite (Hertz and Grindlay 1984)
in the soft X-rays below $\sim$ 4 keV.  
The {\it ROSAT} Galactic Plane Survey (Motch et al.\ 1991) was made as a part of the {\it ROSAT} all sky 
survey, but the energy range was again limited to  below $\sim$ 2 keV.  These soft X-ray surveys
were not able to penetrate the Galactic heavy absorption ($N_H \approx 10^{23}$ cm$^{-2}$),
hence  did not allow  us to observe those X-ray sources located deepest in
the Galactic plane or  behind.

In order to observe completely through the Galactic plane
and to determine the Galactic source population, 
 hard X-ray ($\gtrsim$ 2 keV) observation is  indispensable.
This was made possible with  {\it ASCA}, 
the first imaging satellite in the hard X-ray energy band  (Tanaka, Inoue, \& Holt 1994).
{\it ASCA} carried out systematic surveys  on the Galactic plane
(Sugizaki et al.\ 2001)  and the center region (Sakano et al.\  2002) down to the sensitivity 
limit  $\sim$ 3 $\times 10^{-13}$ erg s$^{-1}$ cm$^{-2}$ in 2 -- 10 keV.
{\it ASCA}'s point source sensitivity was limited
by source confusion due to its moderate X-ray mirror point spread function ($\sim1\arcmin$).
{\it ASCA} discovered more than two hundreds new  X-ray sources on the Galactic plane
region within the longitudes of  $|l| \leq  45^\circ $ (Yamauchi et al.\ 2002; Ebisawa et al.\ 2003).
Many of them  are heavily absorbed and not detected in soft X-ray bands.
{\em ASCA}\/ suggested  an intriguing  possibility that there may be still more dimmer, undetected
 hard X-ray sources in the Galactic plane.
{\em How many Galactic hard X-ray sources are there in the Galactic plane? What is the origin of 
 dimmest Galactic hard X-ray sources?}  Using {\it Chandra}\/, the most sensitive  X-ray 
telescope in the history, we want to answer to these fundamental questions in X-ray astronomy.

Besides the discrete Galactic X-ray sources, 
the Galactic plane itself has been known to emit hard X-rays
 (e.g., Worrall et al.\ 1982; Warwick et al.\ 1985; Koyama et al.\ 1986).
The emission forms a narrow continuous Galactic ridge, thus it is  called
the {\em Galactic Ridge X-ray Emission} (GRXE).
GRXE exhibits emission lines from highly 
ionized heavy elements  such as Si, S and Fe,  which suggests that GRXE originates in
thin hot plasmas with a temperature of several keV (Koyama et al.\ 1986; 
Yamauchi and Koyama 1993; Kaneda et al.\ 1997).
Whether  GRXE is composed of numerous point sources or  truly diffuse emission  has been a big problem
for a long time.  {\it ASCA} was expected to answer this crucial
question, but  not powerful enough 
to clearly separate numerous dim point sources and diffuse emission
 (Yamauchi et al.\ 1996; Kaneda et al.\ 1997; Sugizaki et al.\ 2001).
The origin  of  GRXE remained unresolved with {\it ASCA}.

{\it Chandra} is an ideal satellite that is able to
resolve GRXE into point sources with  a superb  spatial resolution of $\sim 0.''6$
 (Weisskopf et al.\ 2002).
For this purpose,  using the {\it Chandra} Advanced CCD Imaging Spectrometer Imaging array (ACIS-I; Garmire et al.\ 2003),
 we have carried out  deep hard X-ray observations of the
Galactic plane region at  $(l,b) \approx (28.^\circ5,  0.^\circ0)$, where 
extensive observation had been already  made but no discrete X-ray source
detected  with {\it ASCA}.   Our first result was
presented in Ebisawa et al.\ (2001); we have found that 
only $\sim$\mbox{ 10 \%} of the hard X-ray flux (2 -- 10 keV)
in the {\it Chandra} field is explained by the point sources
brighter than $\sim 3 \times 10^{-15}$ erg s$^{-1}$ cm$^{-2}$.
Also, by comparing the observed 
source number density with those measured at high Galactic regions,  
we suggested that most of these hard X-ray point sources on the Galactic plane 
are background AGNs.

We have made two slightly overlapping  {\it Chandra} deep
observations for our project, 
only the first one was analyzed  in  Ebisawa et al.\ (2001).
In the present   paper, we will give full analysis of our two {\it Chandra}\/
observations, as well as the result of near-infrared (NIR) follow-up
observation  using the
New Technology Telescope (NTT) at the European Southern Observatory (ESO).
Main purposes of this paper are the following; (1) to study nature of the dimmest X-ray point sources
on the Galactic plane in details using X-ray and NIR data,  (2) to constrain the population of faint 
Galactic hard X-ray sources, 
and (3) to investigate for  origin of the Galactic diffuse emission
through precise spectral analysis.
Clean separation of the diffuse emission and the
point sources has been made possible for the first time by superb {\it Chandra} spatial resolution.

\section{OBSERVATION AND DATA REDUCTION}
\subsection{X-ray Observation and Data Reduction}
We observed  the ``empty'' Galactic plane region at around 
$(l,b) \approx (+28^\circ.5,0^\circ.0)$ 
in order to study the Galactic diffuse X-ray emission and dim 
point X-ray sources.  This direction is toward  the 
Scutum arm and is known to have strong diffuse emission, thus already
extensively observed with {\it ASCA}
(Yamauchi et al.\ 1996; Kaneda et al.\ 1997).
{\it ASCA} did not detect any  point sources 
brighter than $\sim 2 \times 10^{-13}$ erg s$^{-1}$ cm$^{-2}$ (2 -- 10 keV), 
while  found an intriguing  hard X-ray
diffuse feature (AX J1843.8--0352; Bamba et al.\ 2001), which 
was  another motivation for our {\it Chandra} observation in this region.
Using ACIS-I,
we carried out each 100 ksec pointing on February 25, 2000 (AO1; sequence number 900021) and  
May 20, 2001 (AO2; sequence number 900125), respectively, with slightly overlapping 
fields.
The total area of the observed field is $\sim 250 $ arcmin$^2$. 
Detailed spatial and spectral study of AX J1843.8-0352  using both AO1 and AO2 observations was already
published in Ueno et al.\ (2003).

In this paper, we  use the event data  processed by the {\it Chandra} X-ray Center
with the latest processing system in early 2003. Furthermore, we have applied 
position and energy calibration (CTI correction) 
using the CIAO package (version 3.0.1).  
\anchor{http://www.ipac.caltech.edu/2mass/}{The Two Micron All Sky Survey (2MASS)}\footnote{%
\url{http://www.ipac.caltech.edu/2mass}}
was adopted for the astrometric reference for the  {\it Chandra} data
as well as the NTT/SOFI data (see Section \ref{NIRobs}).
Using the positions of those {\it Chandra} sources that have obvious NIR counterparts,
we estimate our {\it Chandra} position accuracy as $\sim 0.''6$ (see also Section \ref{NIR_id}),
which  is limited  by statistical error and distortion of the
point spread function for off-axis sources. 

\subsection{X-ray  Point Source Extraction}\label{point_source_search}
\begin{figure}
  \begin{center}
    \includegraphics[width=15cm]{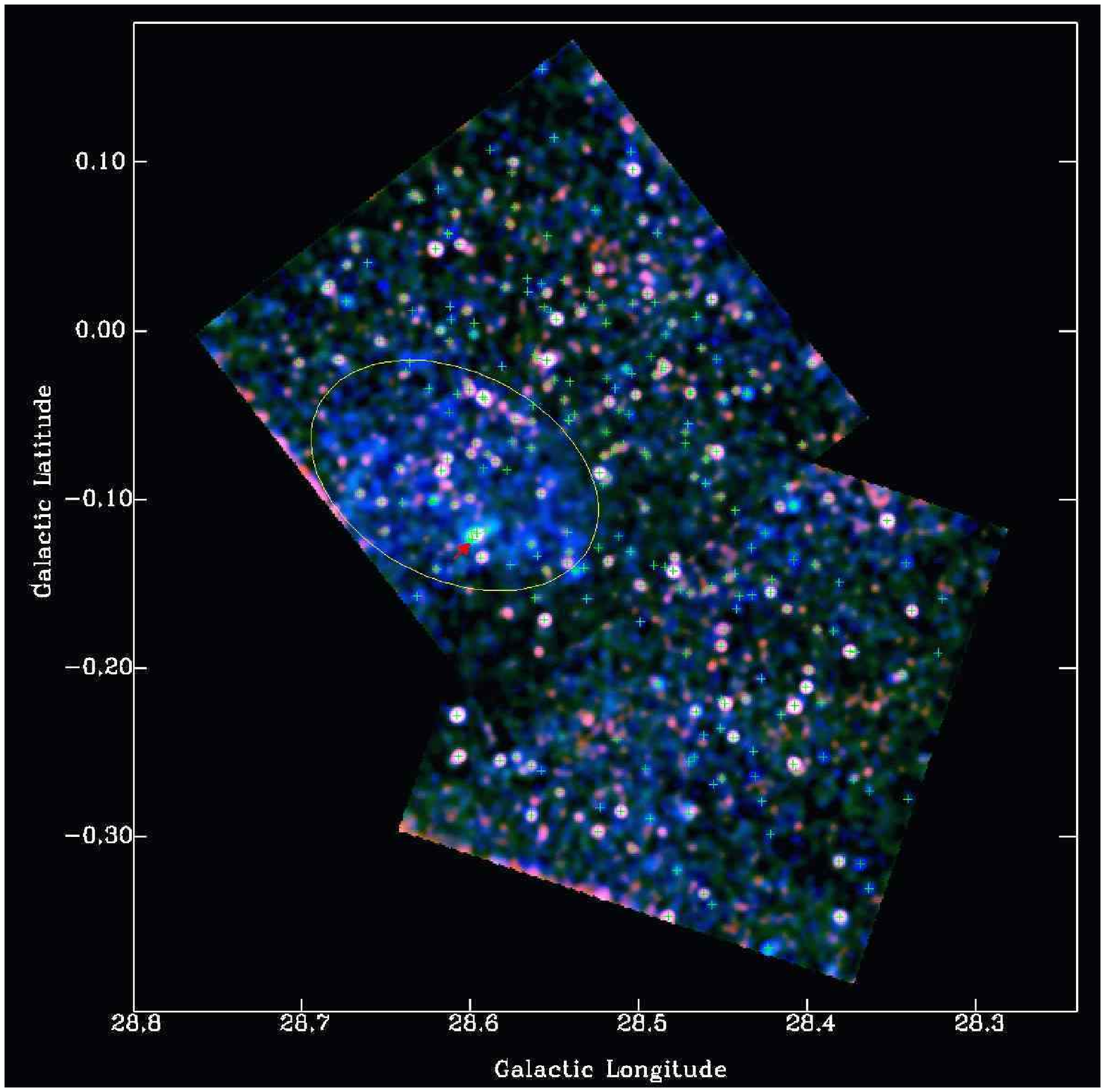}
  \end{center}
\caption{Superposed image of the  two {\it Chandra} observations
with exposure and vignetting correction (in Galactic coordinates).
The upper and lower {\it Chandra} pointing was made in AO1 and AO2, respectively (each 100 ksec).
This is a pseudo-color {\it Chandra}\/ image where  soft X-rays in 0.5 -- 2 keV are represented in red, 
medium X-rays in 2 -- 4 keV in green, and  hard  X-rays in 4 -- 8 keV
in blue.  The image is adaptively smoothed so that both the point sources and the
diffuse emission are clearly  visible.
The 274 detected point sources (Table \ref{srclist_table})  are marked with crosses.
The region including the supernova remnant AX J 1843.8--0352/G28.6--0.1
(Bamba et al.\ 2001; Ueno et al.\ 2003) is shown with a yellow ellipse,
within which the thermal blob structure
CXOU J184357-035441 (Ueno et al.\ 2003) is marked with  a red arrow.
Note that the supernova remnant AX J 1843.8--0352/G28.6--0.1, which exhibits non-thermal
energy spectrum (Bamba et al.\ 2001; Ueno et al.\ 2003),  is more prominent
in hard X-rays (``bluish'' in this representation) than in soft X-rays.
}  
\label{ChandraImage}
\end{figure}

First, we made  exposure and vignetting corrected images in 0.5 -- 2 keV, 
2 -- 4 keV and 4 -- 8 keV, and superposed them by assigning red, green
and blue color respectively, followed by adaptive smoothing to enhance visibility  (Fig.\  \ref{ChandraImage}).
We can clearly see many point sources as well as strong diffuse emission.
We carried out   point source search using the ``wavdetect'' program
in the CIAO data analysis package.  We  searched for  sources in the
0.5 -- 3 keV (soft band), 3 -- 8 keV (hard band)  and 0.5 -- 8 keV (total band) independently.  
We did not use data below 0.5 keV and above 8 keV for the source detection
in order to avoid high backgrounds in both energy ends.
Sources that exceed 4 $\sigma$ significance (given by ``wavdetect'')  in any  of the
three energy bands were considered to be true detections.  
On the AO1 and AO2 overlapping field, we searched for
the sources in  the AO1 and AO2 data separately, and combined the two significances 
 quadratically.
We have  detected  274  point sources  within the total  field of view.
Source position and significance in each energy band 
are given in Table \ref{srclist_table}.  Position errors are  the
statistical ones calculated by wavdetect, not including any systematic errors.

None of these point sources have been reported in X-rays before.
In the soft band, 182 sources have been detected, while in the hard band  
79 sources.  Only 26 sources were detected both in the soft and
hard bands, suggesting an intriguing  dichotomy in the source population (see Sections \ref{logNlogS}
and \ref{NIR_id} for more details).  Those 4 $\sigma$ sources 
near the on-axis roughly corresponds to  $ \sim 3 \times 10^{-15}$  erg s$^{-1}$ cm$^{-2}$ (2 -- 10 keV)
or $ \sim 2 \times 10^{-16}$  erg s$^{-1}$ cm$^{-2}$ (0.5 -- 2 keV), depending on the spectral
hardness ratio and assumed spectral shapes (Section 3.2).  

A new acronym ``CXOGPE''  (Galactic Plane sources reported by Ebisawa et al.\ 2005) is registered at 
\anchor{http://vizier.u-strasbg.fr/viz-bin/DicForm}{CDS}\footnote{\url{http://vizier.u-strasbg.fr/viz-bin/DicForm}}
 for the sources  in Table \ref{srclist_table}.
Hence,  the first source  in Table \ref{srclist_table} at (18:42:51.77, $-3\arcdeg51\arcmin11.\arcsec2$)  may be formally
designated as  ``CXOGPE J184251.7$-$035111'', and so on. In this paper,  however,
the sources in Table \ref{srclist_table} are referred as Source 1, 2, etc. for simplicity.

Sources 208, 210, 213, and 216,  detected at around (R.A., Decl.)$=(18:43:57, -3^\circ54'48'')$ 
seem to be parts of a single extended feature, which is designated as 
CXOU J184357-035441 by Ueno et al.\ (2003).  This extended feature
has a characteristic thermal spectrum (Ueno et al. 2003), and 
is probably a blob associated with  the supernova remnant AX J 1843.8--0352/G28.6--0.1 (Bamba
et al.\ 2001).  All the other sources are consistent with the {\it Chandra}
point spread function, thus they are considered as  point sources.

\subsection{NIR Observation and  NTT/SOFI Data Reduction}

\label{NIRobs}
Because of the heavy Galactic absorption, the NIR band has 
a great advantage over the optical waveband to identify  dim X-ray
sources on the Galactic plane. In fact, radio observation toward
our field has been made (Dickey and Lockman 1990; Dame, Hartman and Thaddeus 2001; Minter
et al. 2001), and the total hydrogen column  density
was estimated not less than $\sim 6 \times 10^{22}$ cm$^{-2}$ (Ebisawa et al.\ 2001).
This corresponds to $A_V \approx 33$ mag
using the standard extinction formula ($N_H/A_V \approx 1.8 \times 10^{21}$ cm$^{-2}$ mag$^{-1}$; Predehl and Schmitt 1995),
so that  there is  hardly a  hope to identify extragalactic sources or 
the most distant Galactic sources in the optical band.
On the other hand, NIR extinction formulas   $N_H/A_J \approx  5.6 \times 10^{21}$ cm$^{-2}$ mag$^{-1}$ (Vuong et al.\ 2003) and
$N_H/E(J-K) \approx   10^{22}$ cm$^{-2}$ mag$^{-1}$ (Harjunpaeae and  Mattila 1996) lead to  $A_J \approx 11$ mag
and $A_K \approx 5$ mag, hence NIR extinction is much smaller, and
NIR observation will allow us  to probe deeper into the Galactic plane.

The 2MASS database covers all the sky, and its Point Source 
Catalog gives the accurate positions and $J$, $H$ and $K_S$ magnitudes 
of all the major stars in our {\em Chandra}\/ field of view (Fig.\  \ref{2mass}).  
In order to study  more deeply than 2MASS,
we have  carried out a NIR follow-up observation 
at ESO using NTT (Tarenghi and Wilson 1989) with the SOFI  infrared camera during 
the nights of 2002 July 28th  and 29th.
SOFI has $1024\times1024$ HgCdTe pixels with a moderately
large field of view ($4.\arcmin94\times 4.\arcmin94$).  The pixel size
is $0.\arcsec2884 $, which is comparable to the  {\it Chandra} one
($\sim 0.\arcsec5$ pixel), thus very suitable to identify {\it Chandra} X-ray sources.
We performed a mosaic observation composed of the 16 pointings slightly-overlapping each other
to cover the central region of the two {\it Chandra} fields  (Fig.\  \ref{SOFI_pointing}).
We defined seven central  ``A'' fields and nine surrounding  ``B'' fields
to cover $\sim$ \mbox{ 75 \%}  of the total {\it Chandra}  field.
The total exposure times  for $J$, $H$ and $K_S$ bands for each A-field are
10,  10, and  14 minutes, respectively, and those for each B-field
are    5,   5, and 7.47 minutes respectively.  According to the
standard SOFI observation procedure, we carried out a dithering
observation, such that each field is covered by 5 ($J$ and $H$ band in B-field), 
8  ($K_S$  band in B-field), 10  ($J$ and $H$ band in A-field) or 15 
($K_S$  band in A-field) sequential ``frames'',  each  jittered by $\sim 40''$.  
Exposure time for each frame was 60 ($J$ and $H$ bands) or 56 seconds ($K_S$ band),
and each frame was furthermore divided into 6 ($J$ and $H$ bands) or 8 
($K_S$ band) read-out segments  in order to avoid saturation.
Thus, each frame was the average of the  6 ($J$ and $H$ bands) or 8 ($K_S$ band) snap-shot images.
The seeing was best in the first night 
($\sim 0.''6$) when we observed fields A1 through  A6 (but only  $H$ and $K_S$ bands for A6), 
whereas in the second night, the seeing was worse ($\sim 1.''5$)
when we observed the remaining fields.







\label{NIR}
IRAF\footnote{IRAF (Image Reduction and Analysis Facility) is distributed by
the National Optical Astronomy Observatories, which are operated by
 Association of Universities for Research in Astronomy, Inc., under
 cooperative agreement with the National Science Foundation.}
 was mainly used to reduce the data following the standard procedure, i.e., 
subtraction of the dark current, flat-fielding using the dome flat, subtraction of sky
using the median-sky technique, removal of bad pixels and cosmic-ray events, and trimming the
frame edges.
SExtractor (Bertin \& Arnouts 1996) was used to extract sources. 
We searched for sources in each  $J$, $H$ and $K_S$ band separately, and detected
16,890,  26,285 and 27,174 sources (with the DETECT\_MINAREA, DETECT\_THRESH, and
ANALYSIS\_THRESH  SExtractor parameter values  5, 1.5, and 1.5, respectively)  
in $J$, $H$ and $K_S$ bands, respectively.  After removing the overlapping sources, 
 32,398 sources have been detected at least in one of the three bands (Table \ref{SOFIcatalog}).

\begin{figure}
  \begin{center}
    \includegraphics[angle=0,width=12cm]{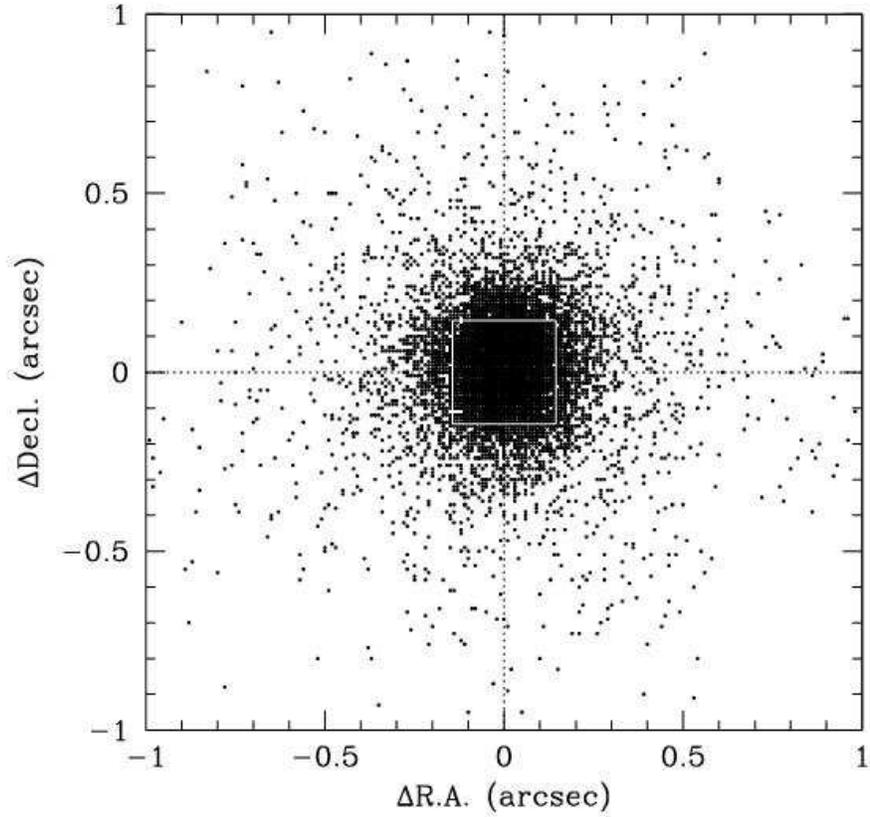}
  \end{center}
\caption{For all the  2MASS sources in the SOFI field of view, relative position of the nearest SOFI source 
(after astrometric correction) is plotted.  
 The central square indicates the SOFI pixel size ($0.''2884\times 0.\arcsec2884$). Standard deviation 
of the positional shift between 2MASS and SOFI  is $0.\arcsec2$ in  R.A. and Decl.}
\label{astrometry}
\end{figure}

Using the 2MASS database as a reference, we have carried out astrometric correction and absolute magnitude
calibration.   In Fig.\  \ref{astrometry}, we compare 2MASS and SOFI positions after the astrometric correction.
The standard deviation of the shifts between 2MASS and SOFI  is $0.\arcsec2$ in  R.A. and Decl.,
which is smaller than the SOFI pixel size ($ 0.''2884$), and may be taken as a typical 2MASS and/or SOFI positional uncertainty.
Considering the 2MASS source within  $ 0.''2884$ radius as a  counterpart,
8,655 SOFI sources are identified in 2MASS (Table \ref{SOFIcatalog}).

\begin{figure}
  \begin{center}
    \includegraphics[angle=0,width=14cm]{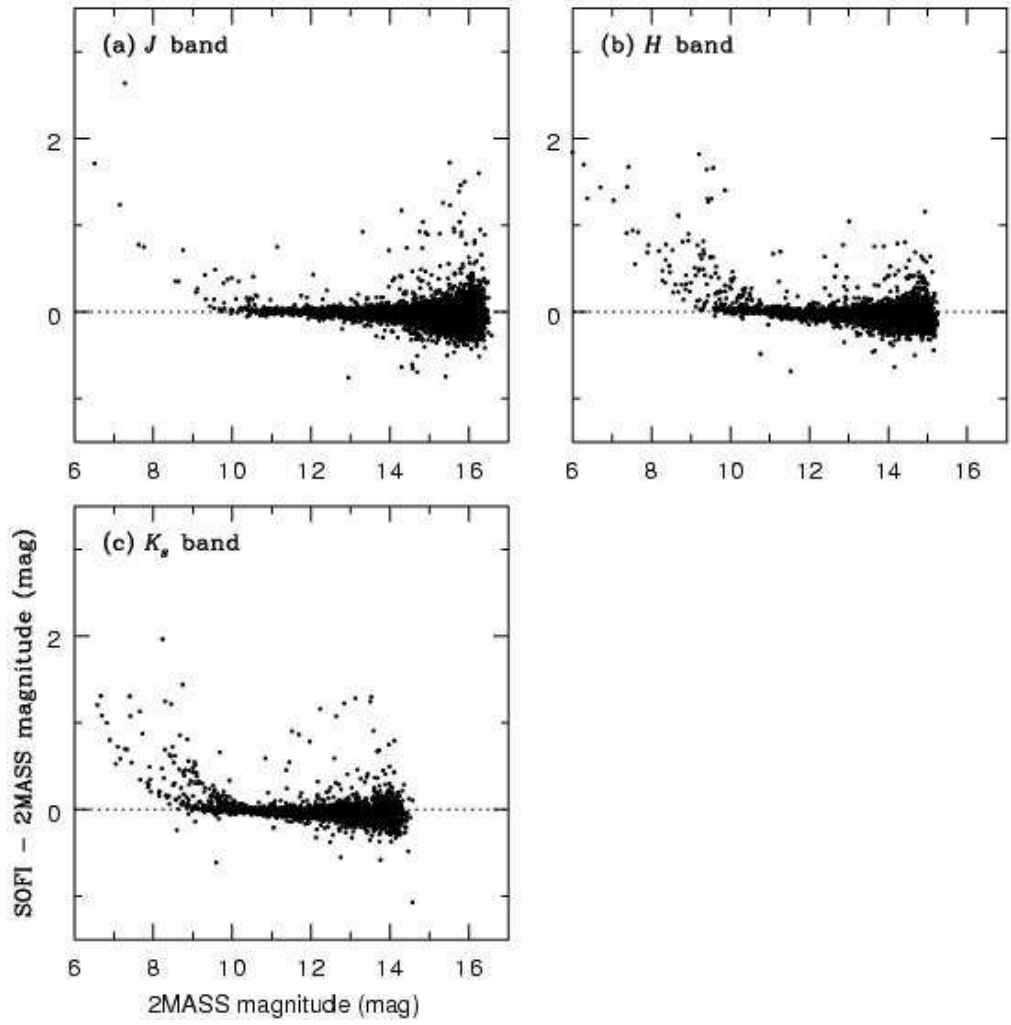}
  \end{center}
\caption{Correlation of the $J$, $H$ and $K_S$ magnitudes of the sources detected by both 2MASS and SOFI.
Difference of the magnitudes is plotted as a function of the 2MASS magnitudes.}
\label{2massSOFIPhoto}
\end{figure}

In Fig.\  \ref{2massSOFIPhoto}
we show correlation of the 2MASS and SOFI $J$, $H$ and $K_S$ magnitudes after the
photometric correction.  
SOFI magnitudes tend to be greater than those of 2MASS for 
sources brighter than $\sim 10$ mag, where SOFI starts to saturate.
Also, there is increasing confusion in the 2MASS-SOFI correlation for stars
fainter than $J\approx$ 14,
$H\approx$13 and $K_S \approx$ 12, because of the relatively large 2MASS pixels.  This leads
to the increasing scatter in the 2MASS-SOFI photometric comparison.
For the cleanest samples between 10 and 12 mag, the standard deviations of 
the 2MASS -- SOFI magnitude differences 
 are 0.08, 0.10, and 0.12 mag in $J$, $H$, and $K_S$ bands, respectively
(for comparison, the 2MASS measurement precision for bright, non-saturated sources 
is $\sim$0.01 to 0.02 mags).  These values may be considered as our typical  photometric uncertainties.

\begin{figure}
  \begin{center}
    \includegraphics[angle=0,width=13cm]{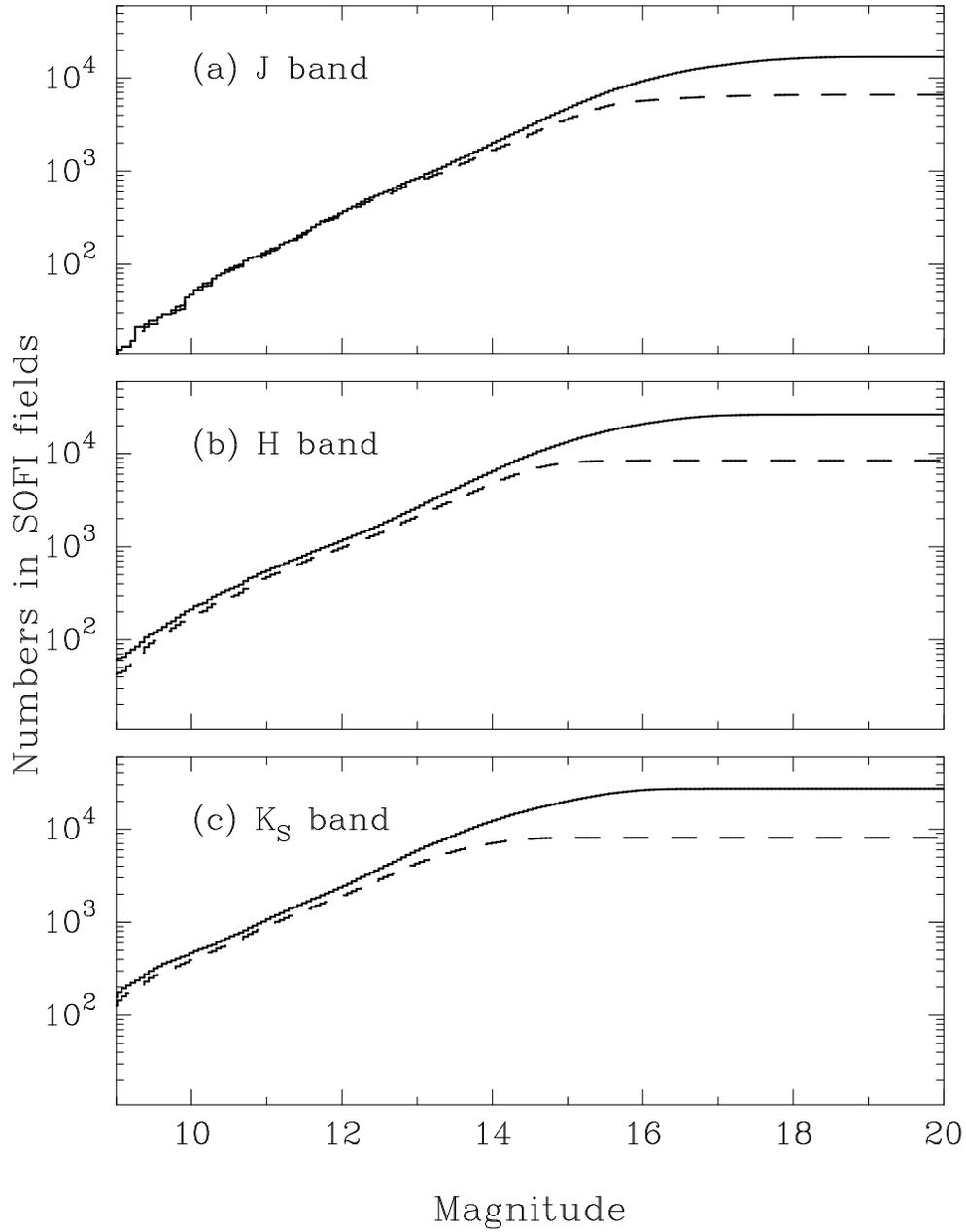}
  \end{center}
\caption{Cumulative histograms of  the number of SOFI 
sources ({\em solid line}\/)  detected in $J$,  $H$ and $K_S$ bands, as a function of the
SOFI magnitudes.  Number of 2MASS sources in the SOFI fields
are also shown with dashed line.
Sources detected in two or three bands are counted in each detected band. 
}
\label{jhk_limmag}
\end{figure}

In Fig.\  \ref{jhk_limmag} we show  cumulative histograms of the SOFI 
and 2MASS sources as a function of the SOFI magnitudes.  We can see that
our sample is almost complete down to $\sim$18, 17 or 16 mag in $J$, $H$, or $K_S$
bands, respectively, if we assume extrapolation of the power-law source 
distribution at the brighter counts.
The depletion of the  sources dimmer than these limiting
magnitudes may be due to our sensitivity limit and/or depletion of the
intrinsic source population.  We can also see the depth of our SOFI
observation relative to 2MASS which saturates at around 
$\sim$16, 15 or 14 mag in $J$, $H$, or $K_S$ bands, respectively.

Finally, we estimate our SOFI sensitivity for normal stars.
We assume an A0 star with the $K$-band absolute magnitude $M(K) = 0$.
At the sensitivity 16 mag in the $K$-band, we would be observing
the A0 star as far as  $\sim 7 $ kpc,  since the source will have
$m(k)= 14.2$ mag without reddening, and $A(K) \approx 1.6$, 
assuming $N_H/A_K \approx 1.3 \times 10^{22} $ cm$^{-2}$ mag$^{-1}$ and 
an interstellar  hydrogen density of $n_H \approx 1$ cm$^{-3}$.

\section{DATA ANALYSIS AND RESULTS}

\subsection{Separation of the Diffuse Emission and Point Sources}\label{seperation}

One of the main goals of the present observation is to  separate the 
diffuse X-ray emission from the point sources,  and to study the diffuse 
spectrum  free from the point source contamination.
We have extracted  an energy spectrum from both AO1 and AO2 data excluding the  AX J 1843.8--0352/G28.6--0.1 region
(marked in Fig.\  \ref{ChandraImage}). 
Using the non-X-ray background database provided by {\it Chandra} X-ray Center (CXC),
we have subtracted  the background spectrum.

\begin{figure}
  \begin{center}
    \includegraphics[angle=0,width=13cm]{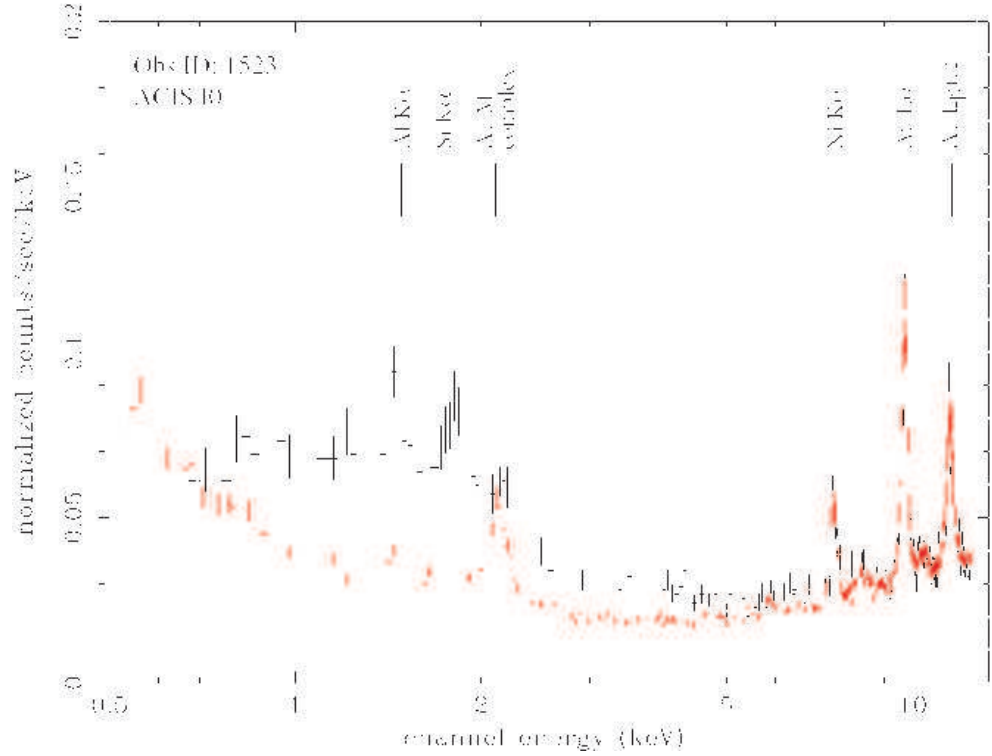}
  \end{center}
\caption{
Histograms of the event count rates of ACIS I0 in our observation ID1523  
({\it black}) and in the CXC background database ({\it red}). The background is normalized in the
energy range 10.5--12.5 keV. Emission lines with the instrumental origin are identified.
}
\label{bgd}
\end{figure}

The CXC background database is constructed from a set of blank sky observations 
at high Galactic latitudes excluding recognized celestial sources. 
We have made background spectrum for each CCD chip, and the relative normalization of the
background subtraction was adjusted for each chip and pointing to cancel the  X-ray
flux in the range of 10.5 -- 12.5 keV (where ACIS is not sensitive to X-rays; Fig.\ \ref{bgd}).
Note that there are several conspicuous instrumental lines in the background spectrum (Fig.\ \ref{bgd};
Townsley et al.\ 2002a,b).
After background was subtracted for each chip and pointing in this manner, 
all the spectra from different chips and pointings have been combined for subsequent spectral study.
After the background is thus subtracted,  our spectrum is expected to include
only point X-ray sources (Galactic and extragalactic) and Galactic plane diffuse emission.
We notice that our results on the Galactic diffuse emission
may not escape  from uncertainty of the background subtraction.
For example, the CXC background might include local soft diffuse emission 
which is presumably higher at high Galactic latitudes.  
However, in the 2 -- 10 keV energy band, which we are most interested in,
the particle background dominates, and its normalization may be correctly
estimated (Fig.\ \ref{bgd}).

\begin{figure}
  \begin{center}
    \includegraphics[width=12cm,angle=270]{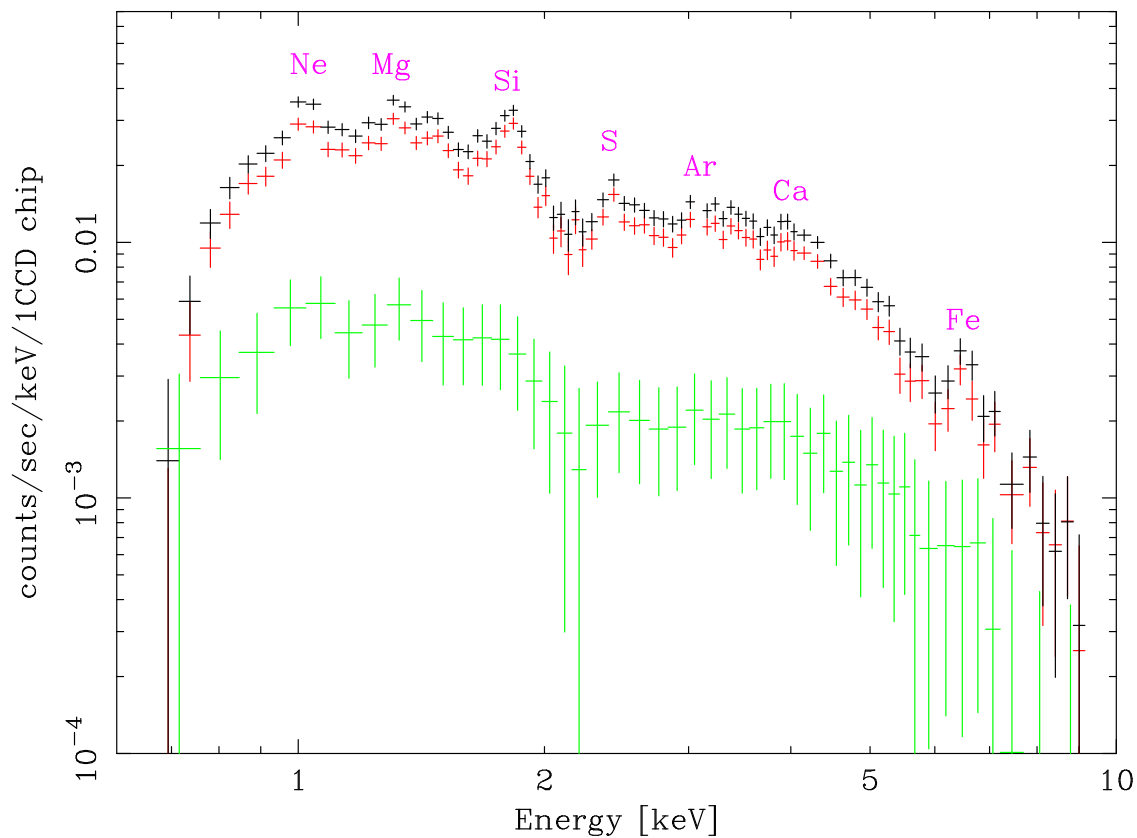}
  \end{center}
\caption{Energy spectra of the total X-rays in the field of view (black;
the AX J 1843.8--0352/G28.6--0.1 region in Fig.\  \protect\ref{ChandraImage} is excluded), of the sum of all the point
sources (green), and of their difference, namely, the Galactic diffuse emission (red).  It is found that $\sim$ \mbox{90 \%} of the X-ray emission
is from the diffuse emission, with which emission lines from highly ionized heavy elements
are associated (prominent emission lines are annotated with element names).  The ordinate is normalized with the average counting rate per CCD chip.}  
\label{diffuse_spec}
\end{figure}

We have made an energy spectrum combining all the point sources in Table \ref{srclist_table}, 
subtracting the background in the same manner. 
In Fig.\  \ref{diffuse_spec}, we show the total energy spectrum and the combined
point source spectrum, as well as the difference  between the two.
We can see  that only $\sim$ \mbox{10 \%} of the total X-ray emission in the {\it Chandra} field of view
is explained by the sum of all the detected point sources.  
Therefore, the difference spectrum, explaining $\sim$ 90 \% of the total flux,
is considered to be mostly the Galactic diffuse emission, though it should include 
some contribution from the point sources below our detection threshold
(see detailed discussion in Section \ref{PointSourceContribution}).
We can see that emission lines from highly ionized heavy elements are associated
with the diffuse emission.  Although presence of the various 
emission lines in GRXE has been  known for a long time (e.g., Koyama et al.\ 1986; Yamauchi and Koyama 1993; Kaneda et al.\ 1997), 
it is now clearly seen that
these emission lines are from the diffuse emission, not from the  point sources.
More detailed diffuse emission spectral analysis  is presented  in Section \ref{diffuse_study}.

\subsection{Spectral Hardness-ratio and Source Fluxes}\label{HRdefinition}

In order to study  spectral characteristics of these new point sources,
we  computed  energy flux and spectral hardness ratio ($HR$) for each source.
The detected raw source counts do not represent the correct source intensities, 
since they are affected by positional dependence of the
detector response (mostly mirror vignetting).
Hence, we define and calculate the ``normalized count rate'' for each energy band, that is 
the count rate  expected when the source is 
located on the ACIS-I aim point (on-axis direction) and observed  for 100 ksec exposure.
We define the spectral hardness ratio as $ HR \equiv (H-S)/(H+S)$, where $H$ is the normalized
count rate in the hard energy band (3 -- 8 keV), and $S$ is that in the soft energy band
(0.5 -- 2 keV).  The normalized count rates and $HR$ are also shown in  Table \ref{srclist_table}.

To determine source energy fluxes, we need to assume  spectral models,
but  most  sources are too dim to determine their energy spectra individually.
Therefore, we took  the following approach: first, energy spectra are extracted 
for all the sources,
and corresponding  instrumental responses  were calculated. 
We categorized  the sources into four spectral groups according to the hardness ratio;
 $HR < -0.8$, $-0.8 \le HR < -0.2$, $-0.2 \le HR <0.6$ and
$0.6 \le HR $. For each spectral group, all the source spectra and responses
were  averaged.  The background spectrum was extracted  in the blank detector field for AO1 and AO2 separately,
and subtracted from the average  spectra (thus, the Galactic diffuse emission is subtracted).
Then, we fitted these four average  spectra with an absorbed power-law model,
and determined the average hydrogen column density and power-law index for each spectral group.
 The best-fit spectral model parameters,
from the softest spectrum  to the  hardest one,
 are  the following : $N_H$ = 0.66, 0.83, 3.4 and  8.0 $\times 10^{22}$ cm$^{-2}$ and photon-index = 4.0, 2.1, 1.9 
and 1.2.  As expected, the higher the $HR$ goes up, the larger the absorption becomes and 
the flatter the spectral slope is.
Finally, each source in the same spectral group was ``fitted'' with the average spectral model  
after background subtraction, 
only by adjusting the normalization.  From the  spectral model thus determined
for each source, we calculated the 0.5 -- 2 keV and 2 -- 10 keV 
energy fluxes (Table \ref{srclist_table}).  
Note,  individual source energy fluxes thus estimated may not  help 
suffering from uncertainty of the spectral assumption.  However,
if we integrate all the sources, the {\em total} point-source energy flux should be valid, 
since we have assumed the average source spectrum for each spectral group.

\subsection{$\log N - \log S$ and Source Population}
\label{logNlogS}

\begin{figure}[ht]
  \begin{center}
     \includegraphics[angle=0,width=13cm]{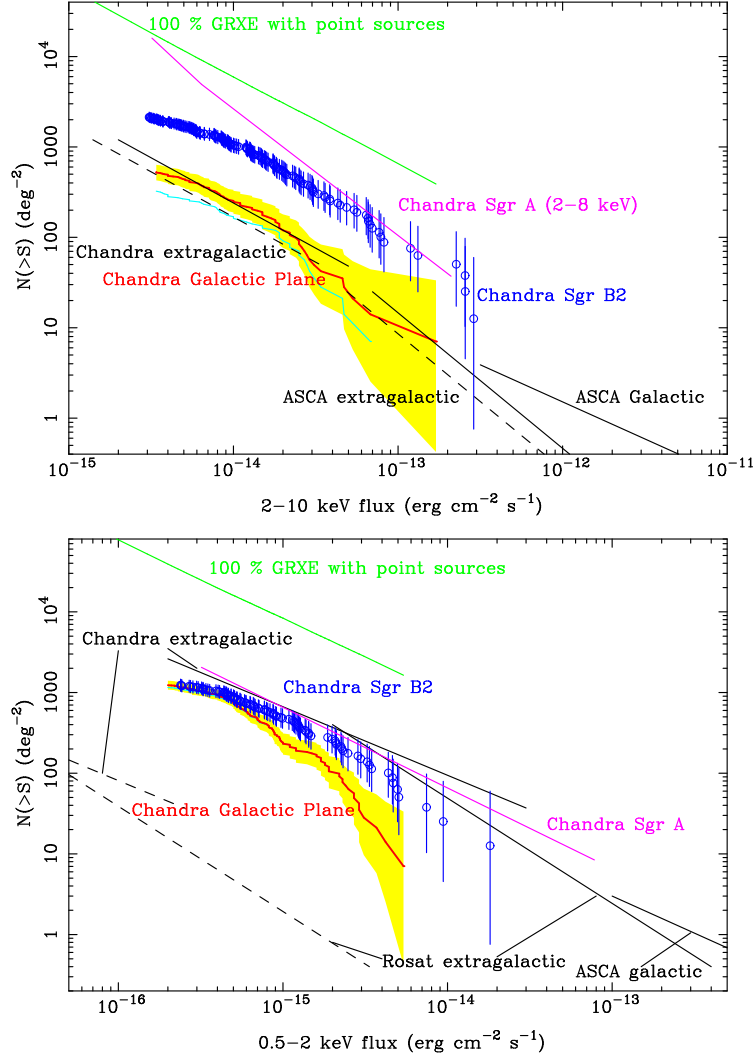}
  \end{center}
\caption{\footnotesize{The $\log N - \log S$ curves of the point sources detected in  our {\it Chandra} Galactic
plane field in 2 -- 10 keV (top) and 0.5 -- 2 keV (bottom). They are indicated in red lines,
and the \mbox{90 \%} error regions are  shown in yellow. 
Also the $\log N - \log S$ curves of  only the sources having the near infrared counterparts 
are shown in cyan (in the soft band, it is almost completely overlapped with red-line,
and barely seen only at the lowest flux).  In addition,   number
of the hypothetical point sources which would account for \mbox{100 \%} of the Galactic ridge X-ray emission
at a given point source flux is indicated in green (see Section \ref{PointSourceContribution} for precise definition).
 Together, other  $\log N - \log S$
relations are shown for the bright {\it ASCA} Galactic sources (Sugizaki et al.\  2001), {\it Chandra} Galactic 
center Sgr B2 and Sgr A (Muno et al.\ 2003),
and extragalactic point sources detected with {\it ASCA} (Ueda et al.\ 1999), {\it ROSAT} and {\it Chandra} 
(Giacconi et al.\ 2001).  For the extragalactic sources, both the original $\log N - \log S$
curves at high Galactic latitudes (black solid lines) and the ones expected on the Galactic plane 
after being extinguished by a hydrogen column density of $6 \times 10^{22}$ cm$^{-2}$ (dashed lines) are shown. 
}}\label{logN-logS}
\end{figure}

We study the point source number densities with  the  $\log N - \log S$ analysis 
in the hard band (2 -- 10 keV) and soft band (0.5 -- 2 keV) separately, adopting
the source  energy fluxes determined above (Fig.\  \ref{logN-logS}).  
The  $\log N - \log S$ curve in the hard band using only the AO1 data was already presented 
in Ebisawa et al.\ (2001).  We set  the lower flux limits at $3 \times 10^{-15}$ erg s$^{-1}$ cm$^{-2}$
in the hard band  and $2 \times 10^{-16}$ erg s$^{-1}$ cm$^{-2}$ in the soft band, respectively,
which approximately correspond to the 4 $\sigma$ detection limits  near the on-axis.


We compare our Galactic plane $\log N - \log S$ curves with those of 
bright {\it ASCA} Galactic sources (Sugizaki et al.\  2001), 
extragalactic point sources detected with {\it ASCA} (Ueda et al.\ 1999), {\it ROSAT} and {\it Chandra} (Giacconi et al.\ 2001).
The extragalactic sources are significantly absorbed on the Galactic plane with a hydrogen
column density of $N_H \approx 6 \times 10^{22}$ cm$^{-2}$ (Section \ref{NIRobs}).
Assuming a typical photon index of $1.7$, we took into account the flux  reduction due to the Galactic absorption,
and made $\log N - \log S$ curves of the extragalactic sources expected to be seen through the Galactic plane
(dashed lines in Fig.\  \ref{logN-logS}).

Let's compare our Galactic plane  $\log N - \log S$ curves with those for extragalactic sources.
In the soft energy band,  number of the
{\it Chandra} sources detected above the lowest flux limit is more than 20 times higher than that expected from the extragalactic sources
through the Galactic plane (dashed line).
Therefore, it is  no doubt that most of the soft sources are Galactic.  On the 
other hand in the hard energy band, the situation is quite different; the extragalactic
$\log N - \log S$ curve explains most of the observed sources on the Galactic plane.
Hence, the observed  dichotomy of the source population 
(Section \ref{point_source_search}) may be naturally explained in that {\em most of the soft X-ray sources 
 have the Galactic origin, while majority of the  hard X-ray sources are extragalactic}.

\subsection{NIR Identification and Source Classification}\label{NIR_id}

\begin{figure}[ht]
  \begin{center}
    \includegraphics[angle=0,width=16cm]{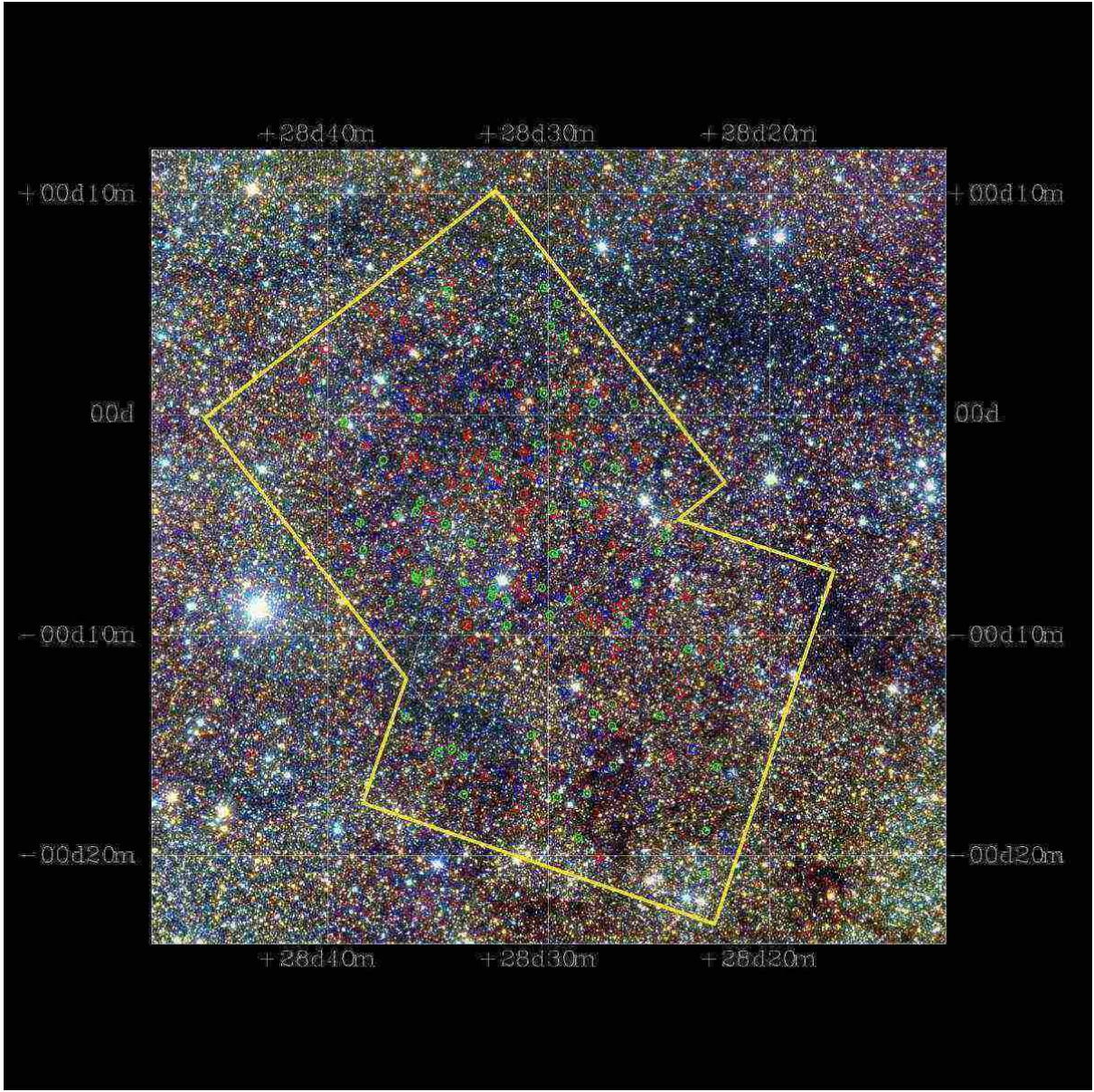}
  \end{center}
\caption{2MASS pseudo-color image of the region including our {\it Chandra}\/ field
(in Galactic coordinates).  The {\it Chandra}\/ field of view is drawn, and the
{\em Chandra}\/ sources are marked in red, green or blue, for soft, medium and
hard sources, respectively.  For the definition of the source spectral
hardness, see Section \ref{HRdefinition}.
}\label{2mass}
\end{figure}

We may classify the new  X-ray sources according to  the $HR$.
We define the ``soft'' sources whose $HR$ are equal or less than the median $HR = -0.60$.
For the rest, the median is $HR =0.11$, so  the ``medium'' and the ``hard'' sources 
are defined as those with $-0.59 < HR \le 0.10$  and
$0.1 < HR$, respectively.  Excluding the sources obviously
associated with the extended structure CXOU J184357-035441 (Ueno et al.\ 2003), number 
of the soft, medium and hard sources we have detected is  136,  65 and 69,  respectively (Table \ref{point_source_table}).
Also, we plot the locations of the soft, medium and hard {\em Chandra}\/ sources on the 2MASS image with different
colors (Fig.\  \ref{2mass}).

Our {\it Chandra} position accuracy is mostly limited by photon statistics and
distortion of the point spread function.  We may expect an error of $\sim 1''$
for dim sources far from the aim points.
We consider the NIR sources found within $\sim 1''$ of the {\it Chandra} positions
as counterparts.  
For the {\it Chandra} sources in the SOFI fields
($\sim$ \mbox{75 \%} of all the {\it Chandra} sources; Table \ref{point_source_table}), 
we show SOFI counterparts in Table \ref{srclist_table}.
2MASS counterparts are given in Table \ref{srclist_table} for 
the {\em Chandra} sources outside of the SOFI fields. 

\begin{figure}[ht]
  \begin{center}
    \includegraphics[angle=0,width=14cm]{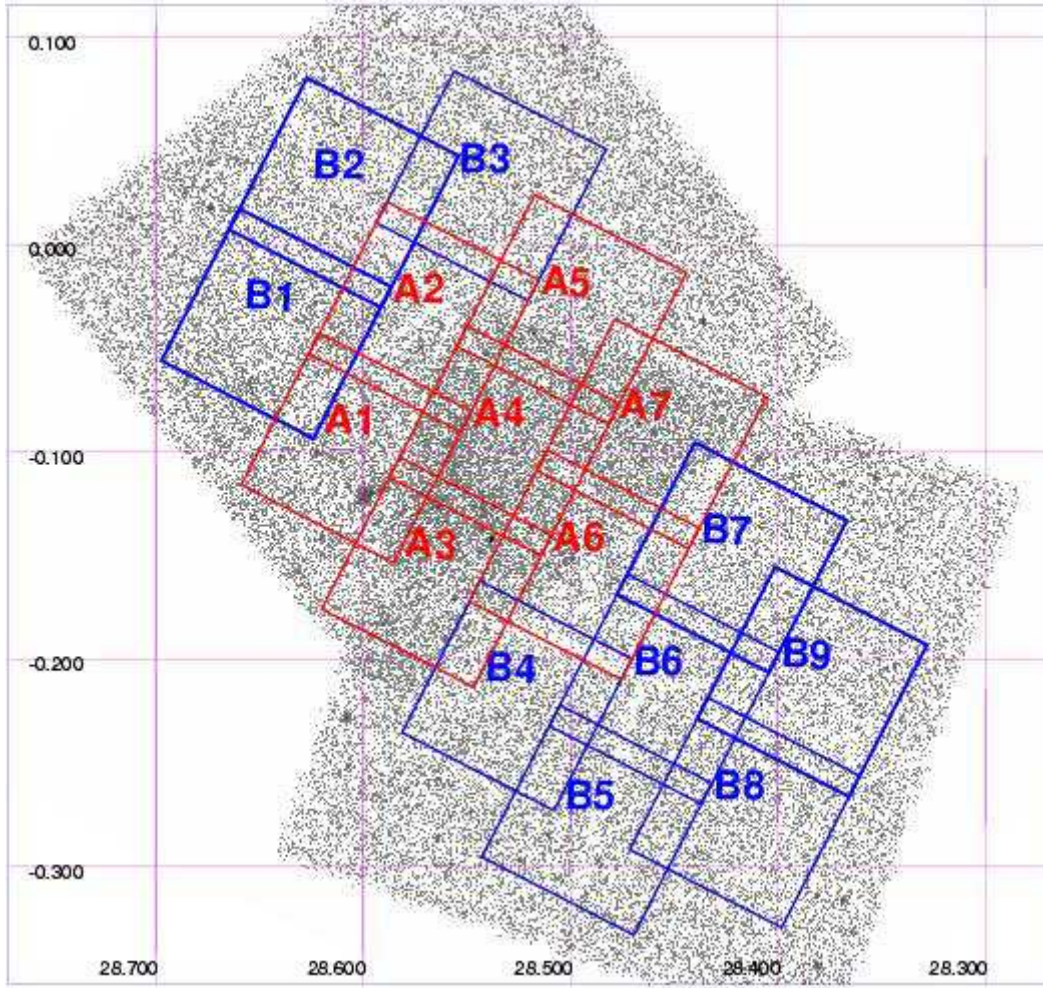}
  \end{center}
\caption{SOFI pointing positions on the {\it Chandra} image (without exposure correction)
 in Galactic coordinates.
Without exposure correction, the AO1 and AO2 overlapping fields and 
CCD  gaps are noticeable (compare with  Fig.\  \ref{ChandraImage}).
}\label{SOFI_pointing}
\end{figure}

The SOFI detected $\sim$ 32,000 sources (Table \ref{SOFIcatalog}) from the $\sim 0.08$ deg$^{2}$
field (Fig. \ref{2mass} and \ref{SOFI_pointing}), namely, the SOFI source
number density is $\sim 0.03$ arcsec$^{-2}$.  Therefore, within $1''$ circle around 
any {\em Chandra}\/ sources, $\sim 0.1$ SOFI source is expected by chance.
In other words, we should be reminded that
 $\sim$10 \% of the {\it Chandra}\/ sources will have accidental SOFI counterparts.

\begin{figure}[ht]
  \begin{center}
    \includegraphics[angle=0,width=14cm]{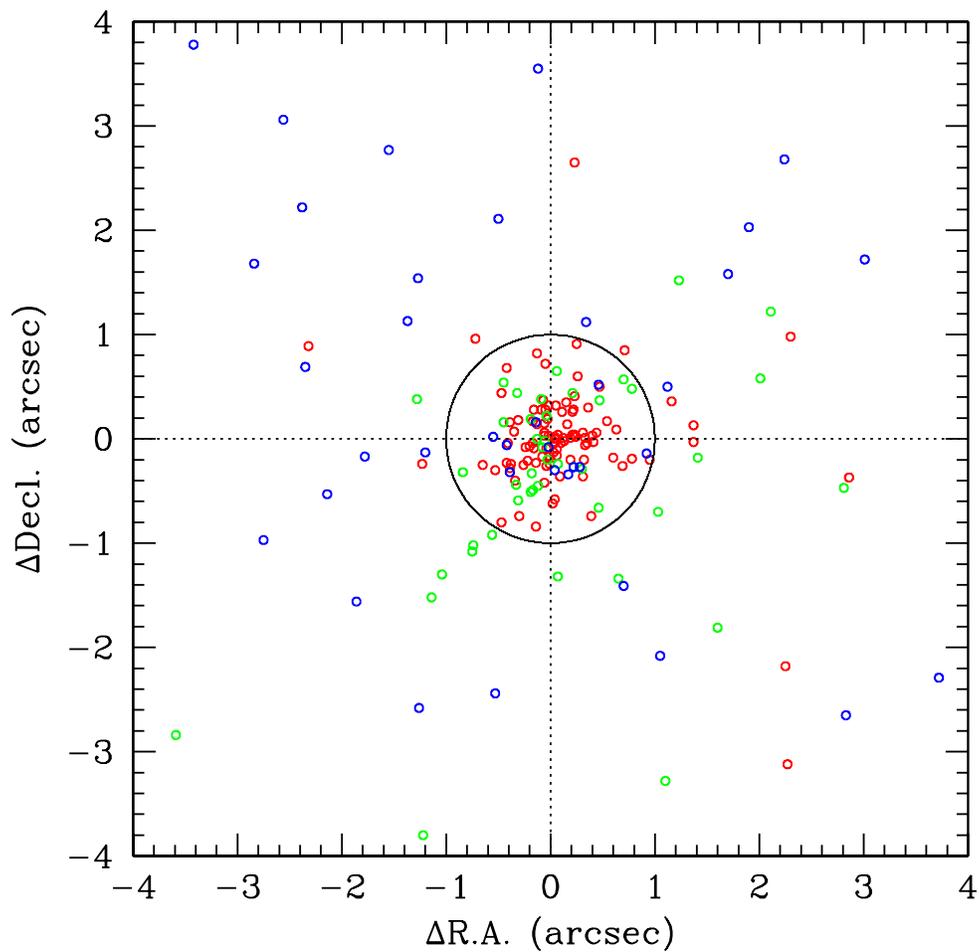}
  \end{center} 
\caption{For all the {\it Chandra} sources detected in the fields covered by SOFI, positional offsets of 
the nearest SOFI sources are plotted.  Red, green and blue are soft, medium
and hard X-ray sources, respectively.  We consider sources within $1\arcsec$  
(central circle)  as the counterparts.  It is obvious that soft X-ray sources
are more likely to have NIR counterparts than harder X-ray sources (see also Table \protect\ref{point_source_table}).
}\label{sofi_chandra}
\end{figure}

\begin{figure}
\centerline{
     \includegraphics[width=14cm]{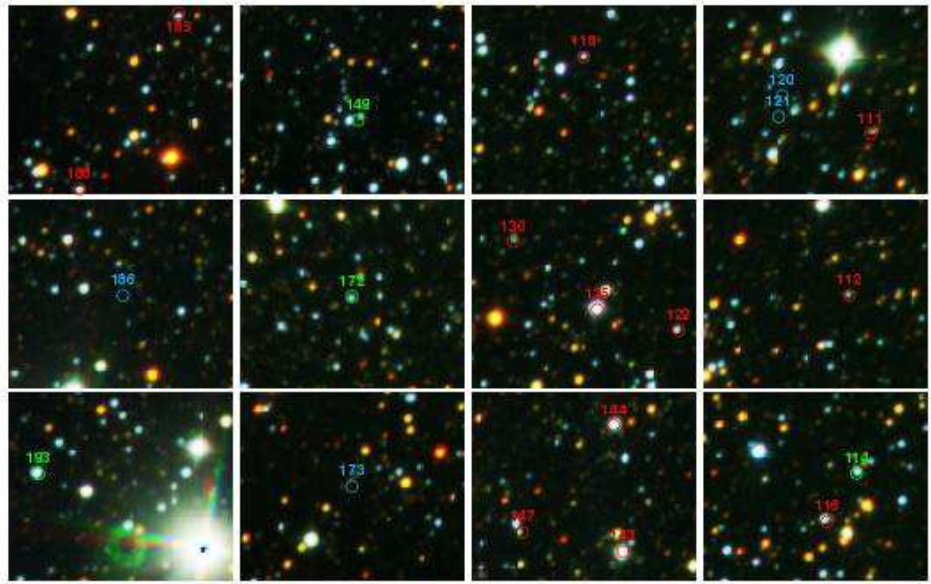}
}
\caption{NIR identification of the {\em Chandra} sources in the
SOFI ``A4'' field (see Fig.\ \protect\ref{SOFI_pointing}).
The pseudo-color image is made by superposing the SOFI $J$, $H$ and $K_S$
band 
images in the RGB scheme.
All the 20 {\em Chandra} sources in this field are marked and labeled with
source ID numbers in Table 1 in red, green and blue for soft,
medium and hard sources, respectively (see section \ref{NIR_id} for 
definition of the spectral hardness).  Radius of each circle is $1''$, 
which is used to identify the NIR counterparts. Note that all the
16 soft and medium sources have NIR counterparts, whereas none of the four
hard sources are identified.
}\label{A4.NIR}
\end{figure}

In Fig.\  \ref{sofi_chandra}, for each of the {\it Chandra} sources 
in the SOFI fields, we plot the
relative positional difference to the nearest SOFI source.
It is easily seen that most of the soft sources
have the NIR counterparts, while medium and hard sources are less likely to
have  counterparts.  In fact, the percentage  of the {\it Chandra} sources having the
SOFI counterparts is  83, 53, and \mbox{22 \%} for soft, medium, and hard
sources, respectively (Table \ref{point_source_table}).
For all the {\it Chandra} sources, the percentage with 
 2MASS counterparts is 
45, 26 and \mbox{12 \%}, respectively  (Table \ref{point_source_table}).
The fact that the detectability of  the NIR counterparts decreases with  $HR$
is  consistent with the idea
that most soft sources are Galactic, while most  hard sources are extragalactic.
In fact, if these hard X-ray sources are background AGNs,
assuming a typical X-ray/NIR luminosity ratio and the significant Galactic extinction, 
they are too dim to be detected in the NIR band (see  Section \ref{origin_point_source}).
Examples of the NIR counterparts for the {\it Chandra} sources are shown in Fig.\ref{A4.NIR}.
This is for the  SOFI ``A4'' field (Fig.\ \ref{SOFI_pointing}), which is covered by both AO1 and AO2
observations.  Note that all the 16 soft and medium sources have NIR counterparts, though
none of the four hard X-ray sources are identified.

\begin{figure}
  \begin{center}
 \includegraphics[height=18cm,angle=0]{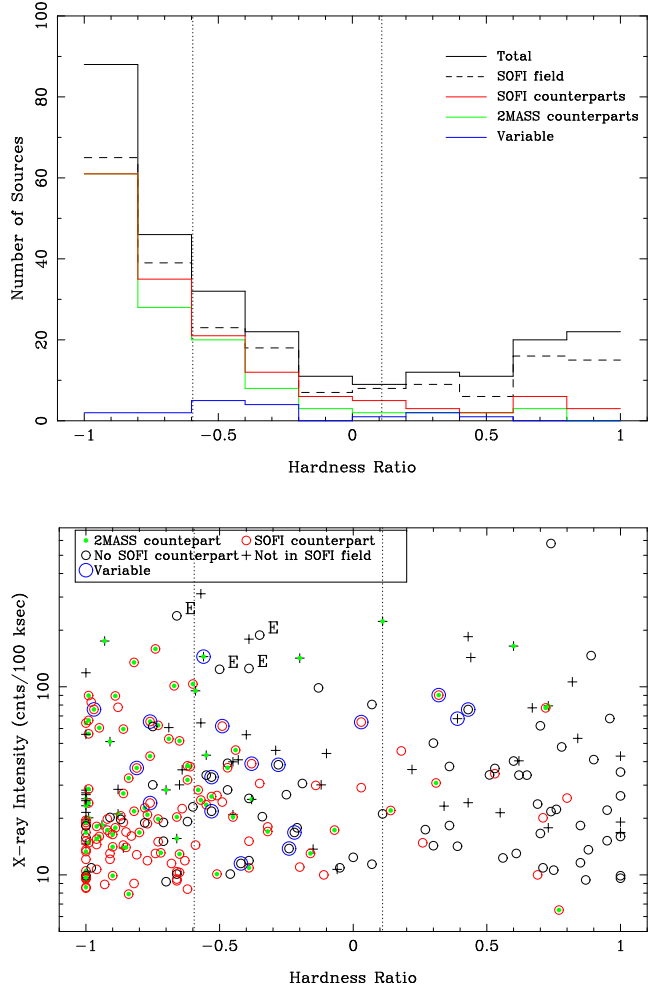}
  \end{center}
\caption{\footnotesize{Histogram of the number of sources as a function of the spectral hardness ratio ($HR$)
(top) and  $HR$  vs.\ the count rate (bottom). In the
top panel, the dashed line indicates the number of sources within the
SOFI fields (Fig.\  \ref{SOFI_pointing}), and the red line shows the sources  having the SOFI 
near-infrared
counterpart.  The green line indicates the number of sources having
 2MASS counterpart, and the blue line indicates the number  of  variable 
sources (Section \ref{variation}). In the bottom panel, the sources outside 
of the SOFI field are shown with crosses, and those inside are with circles: 
black circles indicate the sources without SOFI counterparts, while red circles are
those having  the SOFI counterparts.  In addition, sources having the 2MASS 
counterparts  are marked with green dots, and variable sources
are marked with blue circle. The vertical dotted lines in both figures
indicate the boundaries  we defined between the soft and medium sources ($HR =-0.595$), and
the medium and hard sources ($HR =0.11$).  Source marked with ``$E$''  (Sources 208, 210, 213 and 126) in the bottom panel are parts
of the  extended feature CXOU J184357-035441 (Section \ref{point_source_search}; Ueno et al.\ 2003).
}}\label{hardness}  
\end{figure}

In Fig.\  \ref{hardness}  (top),
we show histograms of the number of sources as a function of $HR$.
It is curious to see that the softest sources are most numerous, and that
the number of sources first decreases with  $HR$ till  $HR \sim 0.5$ and then increases
again.  On the other hand,  number of the sources having 
NIR counterparts decreases monotonically with increasing $HR$.
This also suggests that our source population is composed of the two distinct spectral classes,
the most numerous Galactic soft sources and the less numerous extragalactic hard sources.

The bottom  panel of Fig.\  \ref{hardness} shows the normalized 
X-ray counting rates (Section \ref{HRdefinition}) versus $HR$, as well as 
presence or absence of the NIR counterpart for each source.
We see that almost all the soft sources have NIR counterparts except 
several  dimmest ones.
On the other hand for the hard sources,  presence or absence 
of the NIR counterparts is not related to the X-ray brightness.
Those identified soft X-ray sources are presumably nearby stars,
and the small number of unidentified soft sources are considered to be
farther NIR dim stars (see also Section \ref{PointSourceFitting}). 
The hard X-ray sources without  NIR counterparts are mostly  
likely to be  background AGNs, whose fluxes  are  distributed in a wide range.
In particular, the brightest hard source (Source 200) is not identified in NIR,
thus considered to be a strong  AGN candidate.

\subsection{X-ray Characteristics of the Point Sources}
\subsubsection{X-ray Spectra/Hardness} \label{point_source_spectra}

Since most X-ray  sources are too dim  
(as low as $\sim$10 counts) to make individual
spectra,  we combined the sources having similar spectral hardness and made
average energy spectra to investigate  their spectral characteristics as a class.
We categorized all the point sources  into six groups,
three ranges of the X-ray spectral hardness 
(Section \ref{NIR_id}), and further grouped according to the presence or absence of the NIR counterparts.
We  made an averaged energy spectrum for each group. 

In Table \ref{power_law_fitting}, we show the spectral parameters of the six spectra 
fitted with a simple absorbed power-law model to quantify spectral characteristics. 
For a given spectral hardness, 
the photon index was determined for the group  having 
more sources,  either with or without NIR counterpart, and fixed for the other group. This  helps to clarify the
difference of the hydrogen column densities and normalizations between
the two groups in the same spectral hardness.

Since the hydrogen column density can be  a measure of the distance to the sources,
the gradual increase of the column density from soft to hard suggests that the 
medium and the hard sources are more likely to be located further than
the soft sources. Also, the   medium and  hard sources without NIR counterparts 
show clear excess of the hydrogen column densities compared to those
with the counterparts, which implies 
that the the sources without NIR counterparts
tends to locate further (presumably  extragalactic) than those with NIR counterparts (likely to be Galactic).
Interestingly, in spite that the  sources without NIR counterparts are 
presumably further, they are {\em brighter} in  both the observed fluxes and the intrinsic fluxes
than the sources with NIR counterparts (Table \ref{power_law_fitting}). 
This will make sense that most hard and medium sources are background AGNs which 
are bright in X-rays but not seen NIR.

On the other hand, if we compare the  soft sources without NIR counterparts and those with
counterparts, the average X-ray energy flux of the formers is  $\sim$ \mbox{40 \%}
smaller than the latter (Table \ref{power_law_fitting}).  This suggests that the soft X-ray
sources without NIR counterparts are further than and/or intrinsically dimmer than
those with NIR counterparts.  This makes sense considering that almost all the
soft sources are X-ray active stars in the Galaxy.

\subsubsection{Spectral Fitting}\label{PointSourceFitting}
In the previous section, we used a simple absorbed power-law model
to study the difference of the six average spectra (Table \ref{power_law_fitting}).
Here,  we fit the average  source spectra with more physically meaningful models,
and also study the iron line feature more carefully.

\begin{figure}
\centerline{
 \includegraphics[width=12cm,angle=0]{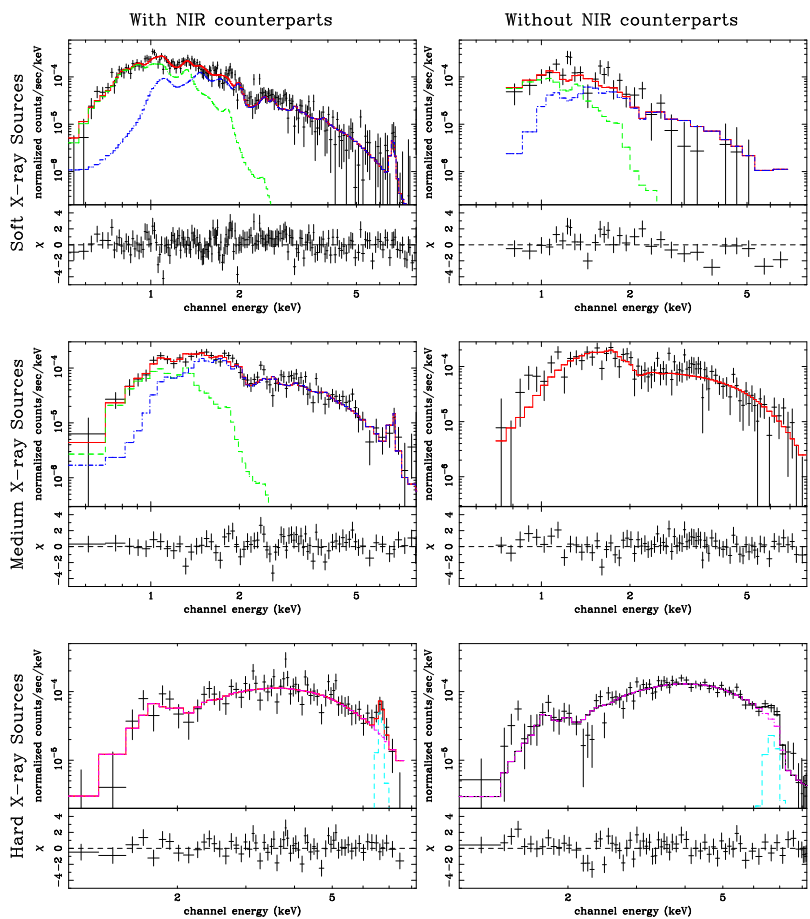}
}
\caption{Composite energy spectra and model fitting 
of the point sources grouped by the X-ray spectral hardness and absence or presence of the NIR counterpart.  Those having the NIR counterpart are in 
the left-hand side, and those without the NIR counterpart are in the right-hand
side.  The top two panels are the soft source spectra, the middle ones are the
medium, and the bottom ones  the hard.
See the text (section \protect\ref{PointSourceFitting}) for the fitting models.
}\label{Souce_Spectra}  
\end{figure}

The X-ray energy spectrum of an active stars is characterized by a  two temperature plasma model.
Therefore, we used a two temperature MEKAL model in XSPEC (version 11.3.1) for the soft spectrum
with NIR counterparts.  The spectrum is well fitted
with a two temperature plasma at  0.2 keV and 2.0 keV   (Table \ref{fitting}).  An ionized iron emission
line is expected from the high temperature plasma, and indeed there is an evidence 
of iron emission line though  not very strong (Fig.\  \ref{Souce_Spectra}, top left).
Assuming that the soft sources without NIR counterpart have the same X-ray spectral properties,
we tried exactly the same  spectral model (including $N_H$) and allowed  only the overall normalization
to be a free parameter.  We found the fit is reasonably well, with \mbox{52 \%} of the
normalization of the soft sources with NIR counterpart (Fig.\  \ref{Souce_Spectra}, top right).  
So, 
soft X-ray sources  without NIR counterparts are dimmer in X-rays than those with NIR 
counterparts due to  intrinsic low X-ray and NIR luminosity and/or due to large distances.

Average energy spectrum of the hard sources with NIR counterparts exhibits a 
conspicuous narrow iron emission line (Fig.\  \ref{Souce_Spectra}, bottom left).
The line center energy is 6.67 keV and the equivalent width is 540 eV (Table \ref{fitting}). 
These iron line parameters and the flat spectrum (photon-index = 1.47) correspond
to a plasma temperature of $kT \sim$ 8 keV.  Such a high temperature thermal spectrum
is a characteristic of quiescent 
cataclysmic variables (e.g., Ezuka and Ishida 1999). In fact, 
cataclysmic variables have been considered prime candidates for faint Galactic hard X-ray sources (e.g.,
Mukai and Shiokawa 1993; Verbant et al. 1997;  Watson 1999).
On the other hand, 
the hard sources without  NIR counterparts do not show a narrow iron emission line, but
have  a broad line and an edge feature which  may be modeled with 
a neutral iron edge (at 7.11 keV) and a broad emission line (at 6.67 keV, EW = 340 eV).
These iron features as well as the flat spectrum (photon-index = 0.77)
are reminiscent of the disk reflection spectrum often seen in Type II AGN.

\begin{figure}[ht]
\plotone{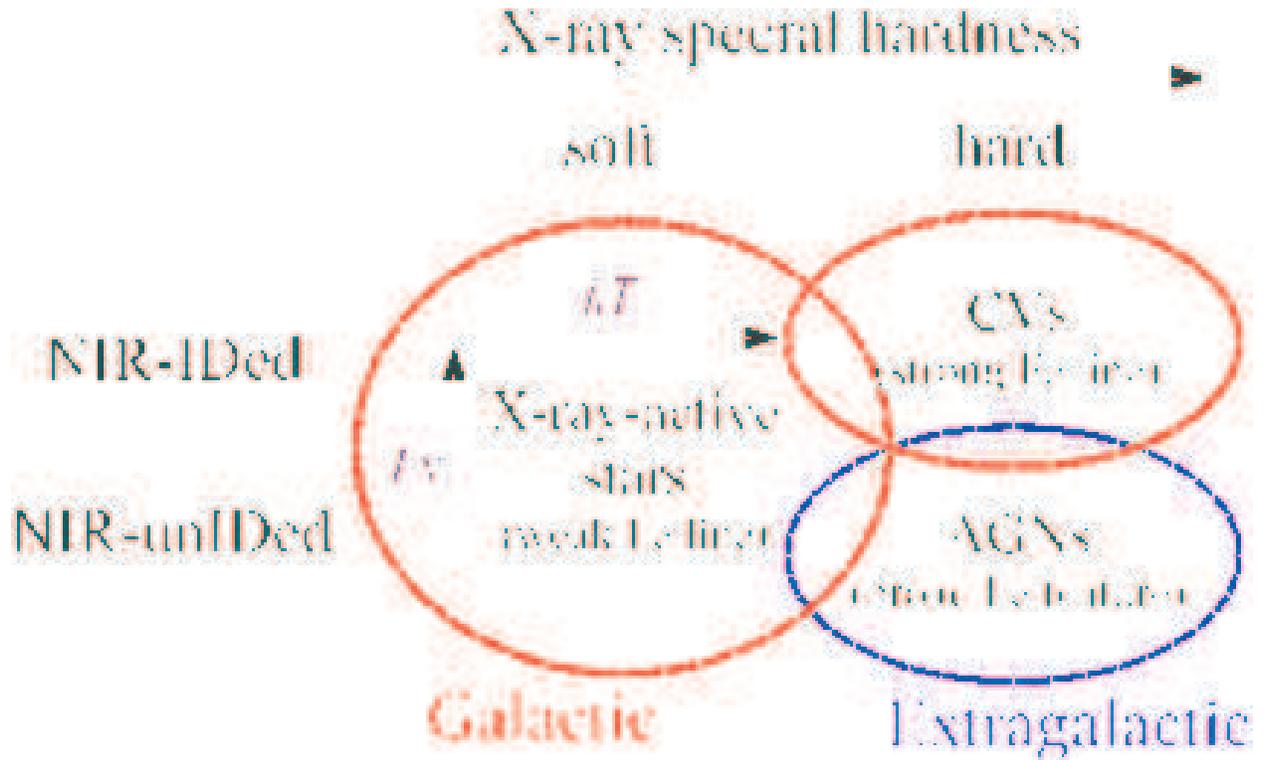}
\caption{Classification of the point X-ray sources according to the X-ray spectral hardness and on the 
presence or absence of the NIR counterpart.}\label{X-ray_NIR}
\end{figure}

The average spectrum of medium sources with NIR counterpart can be fitted with
two temperature plasma model, in which the soft component temperature
(0.2 keV) is similar to that of the soft spectrum but the  hard component
temperature is higher (2.9 keV), suggesting  a
mixture of the relatively hot stars and soft cataclysmic variables.
The average spectrum of medium sources without NIR counterpart can be fitted with a power-law 
which is steeper and less absorbed than the hard sources,
presumably indicating a composite of faint hot stars and  soft AGNs.
A schematic view is shown  in Fig.\  \ref{X-ray_NIR} for  classification of the X-ray sources 
based on X-ray spectral hardness ratio and presence or absence of the NIR counterparts.

\subsubsection{Time Variation}\label{variation}

\begin{figure}[ht]
   \begin{center}
      \includegraphics[angle=0,width=12cm]{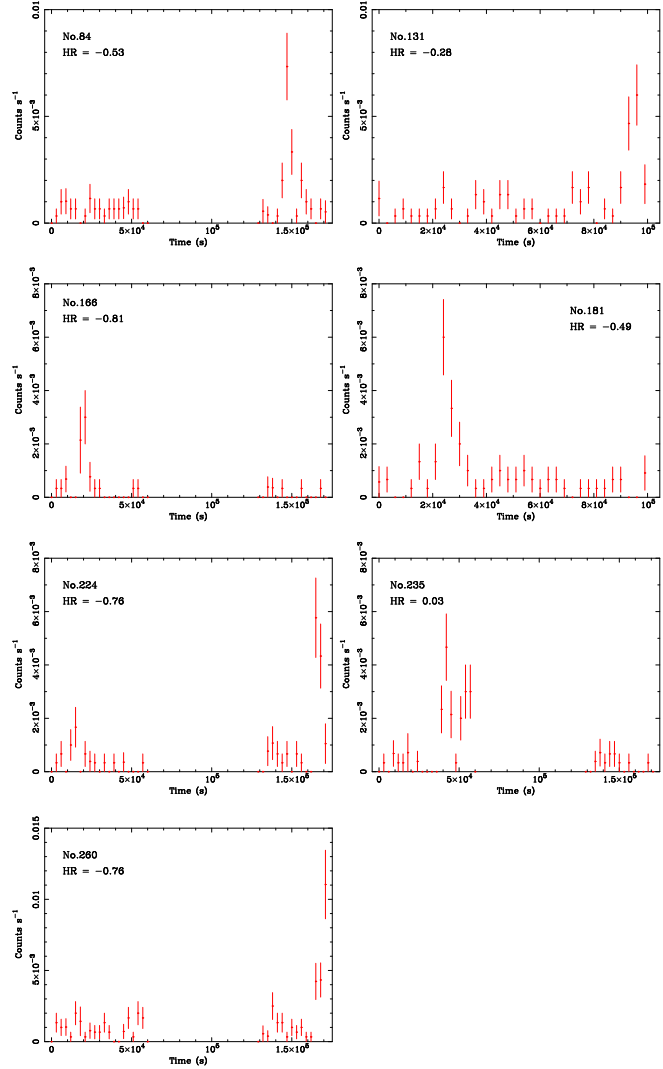}
   \end{center}
\caption{Light curves of the seven variable sources exhibiting  flare-like phenomena.}
\label{lightcurve1}
\end{figure}

We study time variation of the point sources.
For each source, we have made two  light curves
with  bin-widths of 3,000 sec and 10,000 sec.
We performed the Kolmogorov-Smirnov test, and 
if both light curves show variations above   \mbox{99.9 \%} significance level, 
we consider the source to be  significantly variable.
In our sample,  17 sources  are found to be  variable, which are marked in
Table \ref{srclist_table} (``$T$'' in the first column).
The distribution of the hardness ratio for these variable sources is shown
in Fig.\  \ref{hardness} (blue circles).
Flare-like variation of ``rapid-rise and slow-decay'',
which is distinctive to X-ray active stars,  has been found from 
seven sources (Fig.\  \ref{lightcurve1}).  The average $HR$ of these
seven sources is $-0.51$, which reinforces our idea that nature of
these sources is X-ray active stars.

\subsection{Spectral Study of the Diffuse X-ray  Emission}

\label{diffuse_study}



\subsubsection{Line Emission}
First, we concentrate on the iron and other emission lines of the diffuse emission.  We fit the
iron energy band (5.5 -- 7.2 keV) and soft energy band (0.8 -- 3.5 keV) separately with a
simple power-law plus gaussian model, and determine the line parameters.

\begin{figure}[ht]
    \begin{center}
      \includegraphics[angle=0,width=12cm]{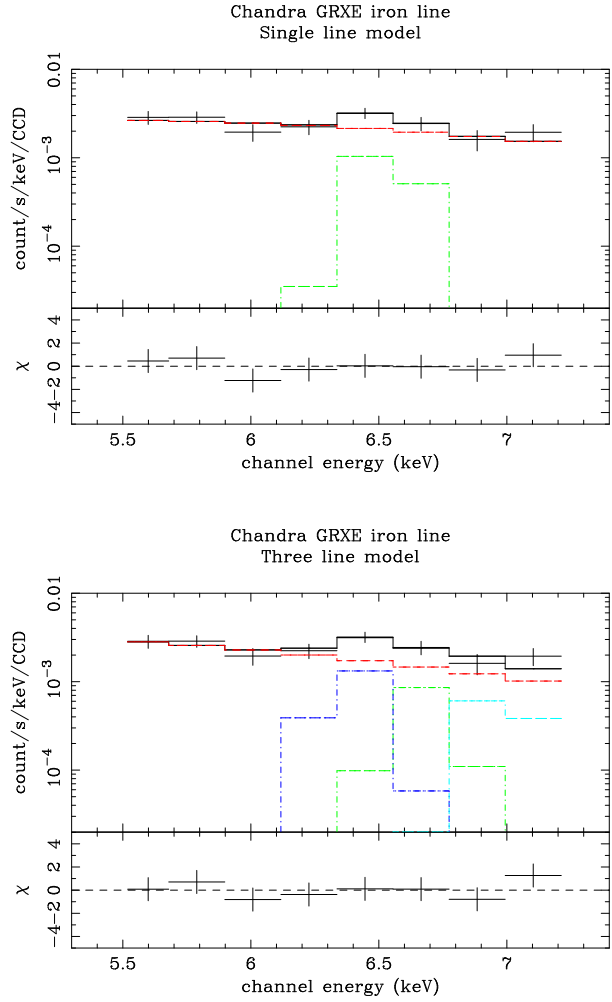}
   \end{center}
\caption{The iron line from the Galactic ridge diffuse X-ray emission
fitted with a single line (top) or three lines (bottom; differentiated with different colors).
Both models fit the observed spectrum  equally  well.
 }
\label{iron_line}
\end{figure}

We point out that ACIS-I iron line measurement has
a significant merit in that contaminating instrumental iron line, which was  problematic in {\it ASCA} and 
{\it XMM-Newton} diffuse spectral  study,  is almost fully negligible.
In the iron energy band, a single narrow gaussian model is
successful (Table \ref{iron_line_fitting}; Fig.\  \ref{iron_line}, top).
The central line energy is $6.52^{+0.08}_{-0.14}$ keV (\mbox{90 \%} error), which is
consistent with Kaneda et al.\ (1997; $6.61\pm0.02$ keV), and significantly lower than
what expected from He-like iron in a thermally equilibrium plasma (6.67 keV).
A possible explanation of the line energy shift is that the plasma is in non-equilibrium ionization
(NEI) state (Yamauchi and Koyama 1993; Kaneda et al.\ 1997), or the line is composed of a
fluorescent 6.4 keV line and a thermal 6.67 keV line (Valinia et al.\ 2000b).
Considering the latter possibility, we fit the same spectrum with a multiple line model.
In addition to these two lines, if the charge exchange takes place between the
cosmic-ray iron nuclei and interstellar hydrogen atoms
(Tanaka, Miyaji, \& Hasinger 1999; Tanaka 2002), a hydrogenic iron line at 6.97 keV is expected.
  Thus, we fit  the observed spectrum with the three lines with fixed energies
(Table \ref{iron_line_fitting}; Fig.\  \ref{iron_line}, bottom).
The fit is acceptable, though slightly worse than with the single line model.
The cosmic-ray charge exchange model predicts significantly broadened emission lines
due to the energetic cosmic-ray bulk motion (Tanaka, Miyaji, \& Hasinger 1999; Tanaka 2002), but
from our statistics we could not constrain the intrinsic iron line width.

We found  that the iron line equivalent width in GRXE
is significantly dependent on the spectral model, and difficult to be determined uniquely.
With a single narrow line model we obtained $EW = 170\pm120$ eV, 
which is smaller than the {\it ASCA} value with the same model and same sky region,  $405\pm80$ eV 
(Kaneda et al.\ 1997).  Contamination of the point sources with strong iron emission in the {\it ASCA} spectrum
might explain the different at least to some extent.
On the other hand, with the three line model, our equivalent width values 
are $100\pm^{50}_{100}$ eV, $180\pm^{360}_{140}$, and  $160\pm^{260}_{160}$ eV,
which are consistent with the {\it ASCA} result using the same three line model on the same sky,
$<70$ eV, $280\pm70$ and $120\pm70$ eV (Tanaka 2002), though errors are large.

\begin{figure}[ht]
   \begin{center}
     \includegraphics[angle=270,width=12cm]{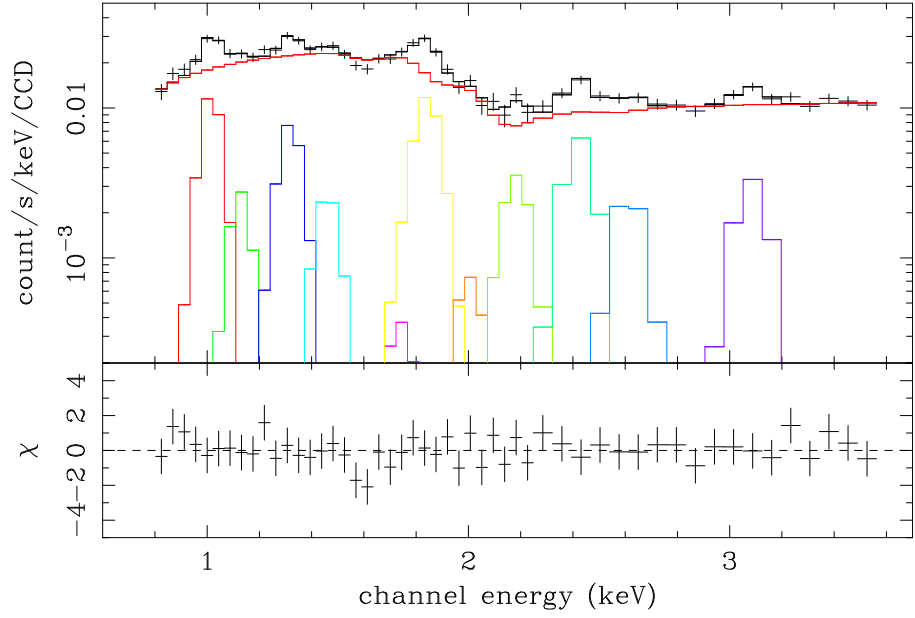}
   \end{center}
\caption{Emission lines from the Galactic ridge diffuse X-ray emission in the soft energy band.
Eleven gaussians (differentiated with different colors) are put with a power-law continuum.
 }
\label{soft_line}
\end{figure}

Similarly, the soft band energy spectrum was fitted with a power-law continuum and eleven gaussians
(Table \ref{soft_line_fitting}; Fig.\  \ref{soft_line}).  These are the same lines
detected in {\it ASCA} (Kaneda et al.\ 1997),  with an additional line at  2.19 keV which  probably
originates in the instrumental Au M complex (Fig.\ \ref{bgd}).
Equivalent width values are consistent with those in Table 4 in Kaneda et al.\ (1997),
except that we find weaker 1.74 keV (low ionized Si) and 2.00 keV (\ion{Si}{14},Ly$\alpha$) lines.

\subsubsection{Fit with Non-Equilibrium Ionization Plasma Model}
\label{Fit_with_NEI}
We now try to fit the observed diffuse spectrum with a more physically reasonable model.
As a working hypothesis, we adopt the same spectral model 
used by Kaneda et al.\ (1997), which is a two temperature NEI model
(Masai 1984), such that there are soft and hard NEI components which have
different temperatures, normalizations, ionization parameters and are affected by 
different amounts of interstellar absorption.

\begin{figure}[ht]
   \begin{center}
      \includegraphics[angle=0,width=12cm]{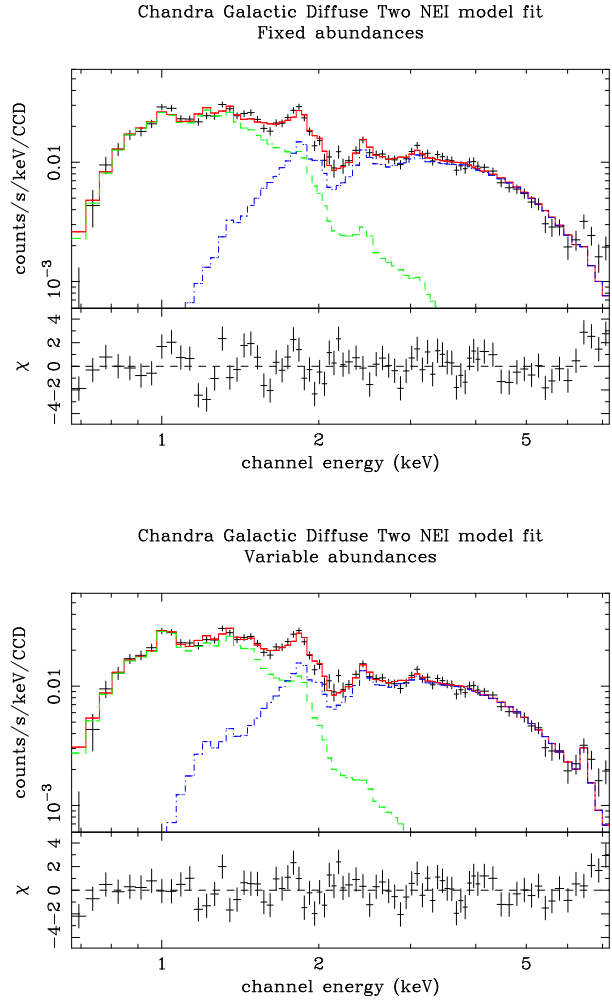}
   \end{center}
\caption{Spectral fitting of the Galactic diffuse emission
with a two component NEI model (Masai 1984) with fixed abundances for 
each component (top) or variable abundances (bottom).
Note that the iron line ($\sim$ 6.7 keV) and neon line ($\sim$ 1.0 keV) 
features are 
fitted in the bottom figure, but not in the top.
 }
\label{GRXE_fit}
\end{figure}

First, we fix the element abundances for the soft component and
hard component, respectively (Table \ref{NEI_fitting} left; Fig.\  \ref{GRXE_fit} top).
The fit is not satisfactory (reduced $\chi^2 = 1.97$), and in particular, the observed iron and neon emission lines are not
explained.  Next, we adjust abundances of Ne, Mg, and Si in the soft component,
and Fe abundance in the hard component (Table \ref{NEI_fitting} right; Fig.\  \ref{GRXE_fit} bottom).
Now the fit is better (reduced $\chi^2 = 1.52$), though artificial adjustment of the abundances is unexplained.
Still, we notice a hint of high energy excess above iron line energy, which may be related to the
non-thermal component reported above $\sim$ 10 keV (Yamasaki et al.\ 1997; Valinia et al.\ 2000b).

\begin{figure}[ht]
   \begin{center}
     \includegraphics[angle=270,width=12cm]{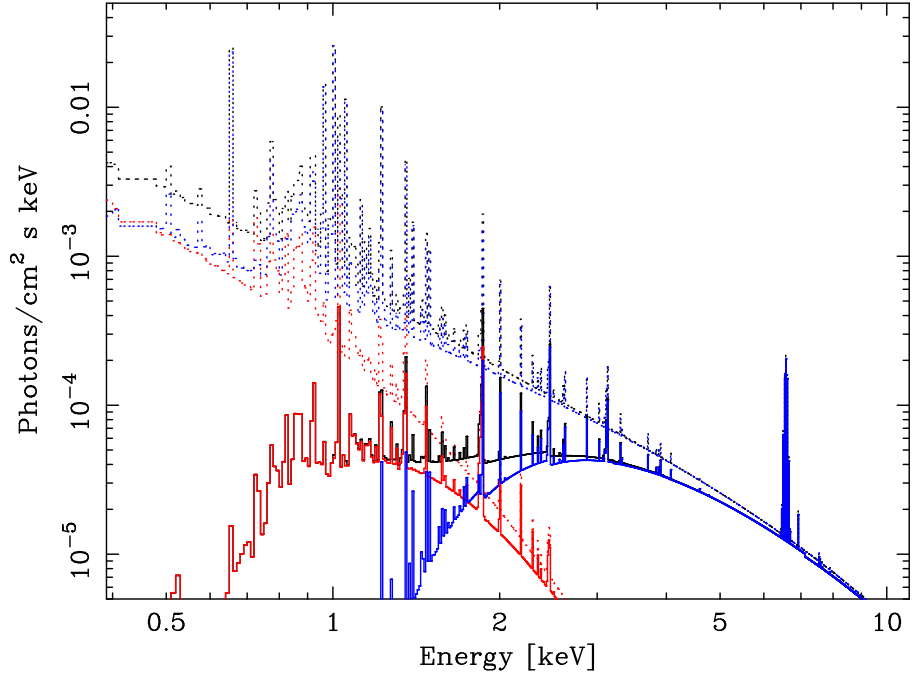}
   \end{center}
\caption{Best fit spectral model used in Fig.\  \protect\ref{GRXE_fit}
including interstellar absorption (solid lines) and removing the  absorption (dotted lines).
Hard component is drawn in blue, soft component is in red, and total is in 
black.
 }
\label{GRXE_model}
\end{figure}

With the two component model fit, we have determined the observed flux
from the soft and hard component as 
$2.6 \times 10^{-8}$ and  $2.1 \times 10^{-7}$ erg cm$^{-2}$ s$^{-1}$ str$^{-1}$ (0.7 -- 10 keV), respectively.
Note that the observed flux and spectral shape are significantly
affected by the heavy interstellar absorption (Fig.\  \ref{GRXE_model}). 
Although the soft component is dominant in the observed flux below 2 keV, the
hard component is more dominant over the entire energy band if absorption is removed.
If the absorption is removed, the intrinsic fluxes are 
$1.1 \times 10^{-7}$ and  $7.6 \times 10^{-7}$ erg cm$^{-2}$ s$^{-1}$ str$^{-1}$ (0.7 -- 10 keV) from the
soft and hard component,  respectively\footnote{Using the same two component model, Kaneda et al. (1997) gave
the intrinsic (= absorption removed) soft and hard component fluxes 
$1.9 \times 10^{-6}$ and  $5.3 \times 10^{-7}$ erg cm$^{-2}$ s$^{-1}$ str$^{-1}$ (0.5 -- 10 keV), respectively.
We remark that,
in the NEI models we assume,  strong oxygen lines are expected between 0.5 keV and 0.7 keV (Fig.\  \ref{GRXE_model}),
which are hardly observable  below the low energy thresholds of both  {\it ASCA} and {\it Chandra}.  
Therefore, it will be more reasonable to compare 
the  fluxes in 0.7 -- 10 keV, not in 0.5 -- 10 keV.
Calculated from the NEI model parameters by Kaneda et al. (1997), 
 the observed soft and hard component fluxes with {\it ASCA} are
 $5.3 \times 10^{-8}$ and  $1.2 \times 10^{-7}$ erg cm$^{-2}$ s$^{-1}$ str$^{-1}$ (0.7 -- 10 keV), respectively, and 
the intrinsic soft and hard component fluxes  are
$1.9 \times 10^{-7}$ and  $4.3 \times 10^{-7}$ erg cm$^{-2}$ s$^{-1}$ str$^{-1}$ (0.7 -- 10 keV), respectively.
Note that excluding  the 0.5 -- 0.7 keV energy range makes the intrinsic soft component flux  10 times less.
Flux difference between {\em ASCA} and {\em Chandra} is discussed in Section \ref{comparison}.}.

\section{Discussion}
\subsection{Origin of the Faint X-ray Point Sources on the Galactic Plane}
\label{origin_point_source}
We have detected 270 point X-ray sources (above 4 $\sigma$ significance)
on a typical Galactic plane field  at around $(l,b) \approx (28.^\circ5,  0.^\circ0)$
within $\sim $250 arcmin$^2$, down to
the flux limits   $\sim3 \times 10^{-15} $  erg s$^{-1}$ cm$^{-2}$ (2 -- 10 keV) or
$\sim2 \times 10^{-16} $ erg s$^{-1}$ cm$^{-2}$  (0.5 -- 2 keV).  Thereby  we
extended the X-ray $\log N - \log S$ curves on the Galactic plane 
to much dimmer levels than previous observations.
In the brightest ends, our $\log N - \log S$ curves match well
with those by {\it XMM-Newton} (Hands et al.\ 2004) and {\it ASCA} (Sugizaki et al.\ 2001)
carried out on larger Galactic plane regions. 

Based on the X-ray spectral properties and presence  or absence of the NIR counterparts,
we have proposed a schematic view for  origin of the point X-ray sources on the
Galactic plane;  soft X-ray sources  are nearby active stars,  whereas
hard X-ray sources without NIR counterpart are extragalactic and
hard sources with NIR counterpart are Galactic cataclysmic variables
(Fig.\  \ref{X-ray_NIR}).   
In the following, we are going to confirm this simple picture  through more detailed analysis.

Considering background AGNs,  assuming typical X-ray fluxes (Table \ref{power_law_fitting})
and  a  broad-band photon-index of 2 (flat energy spectrum in the $\nu F\nu$ plot),
we estimate the expected $J$, $H$ and $K_S$ magnitudes  as 21 mag to 23 mag.
Taking  into account the further reddening $A_K\approx 5$ (Section \ref{NIRobs}), 
there is  no hope to detect those background AGNs in NIR through the Galactic plane.
Therefore, we may assume that all the  sources with NIR identification  are Galactic.

\begin{figure}[ht]
   \begin{center}
     \includegraphics[angle=0,width=12cm]{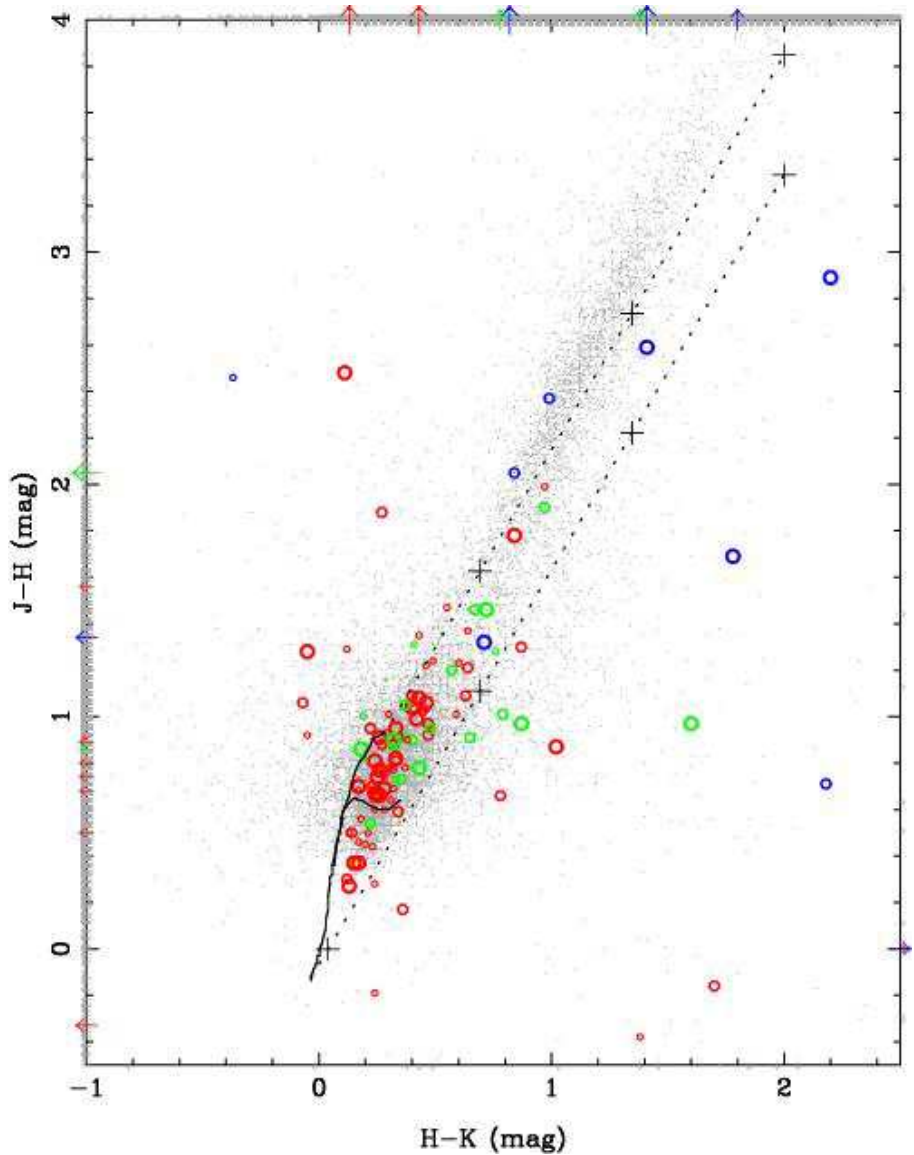}
   \end{center}
\caption{NIR color-color diagram of all the SOFI sources
(gray) and the {\it Chandra} sources with NIR counterparts (red, green and  blue circles 
for soft, medium and hard sources, respectively). The symbol size for {\it Chandra} sources
is approximately proportional to the X-ray counting rates.
Solid curves indicate the loci for dwarfs (main sequence stars) and giants,
and dotted lines show the reddening (towards upper-right), such that distance between the crosses
correspond to $A_V$=10 mag  (data taken from Cox 1999). Sources detected only in two bands
are marked at the border of the graph.  Note that there
are many SOFI sources which are detected only in one band,
thus cannot be  plotted in this diagram.
 }
\label{color_color}
\end{figure}

\begin{figure}[ht]
   \begin{center}
      \includegraphics[angle=0,width=13cm]{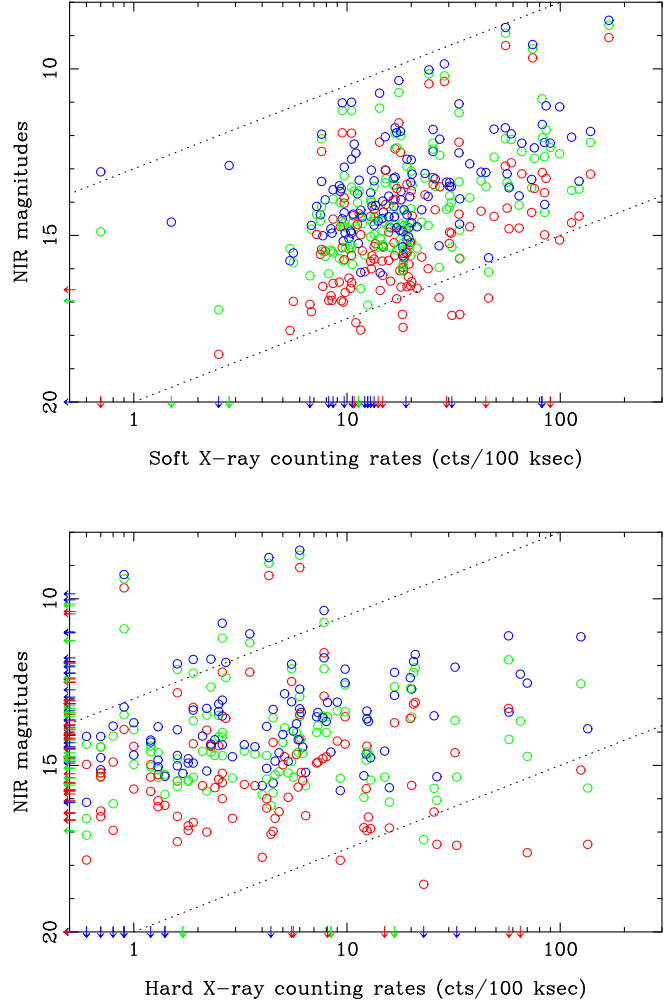}
   \end{center}
\caption{Correlation between  the 
X-ray counting rates  (top = 0.5 -- 2 keV  band, bottom= 2 -- 10 keV band) and
the NIR magnitudes  (red=$J$, green =$H$ and  blue = $K_S$).
The black dotted lines  indicate the slopes for the
sources with the constant luminosities at different distances.
 }
\label{X-ray_NIR_corr}
\end{figure}

\begin{figure}[ht]
   \begin{center}
      \includegraphics[angle=0,width=13cm]{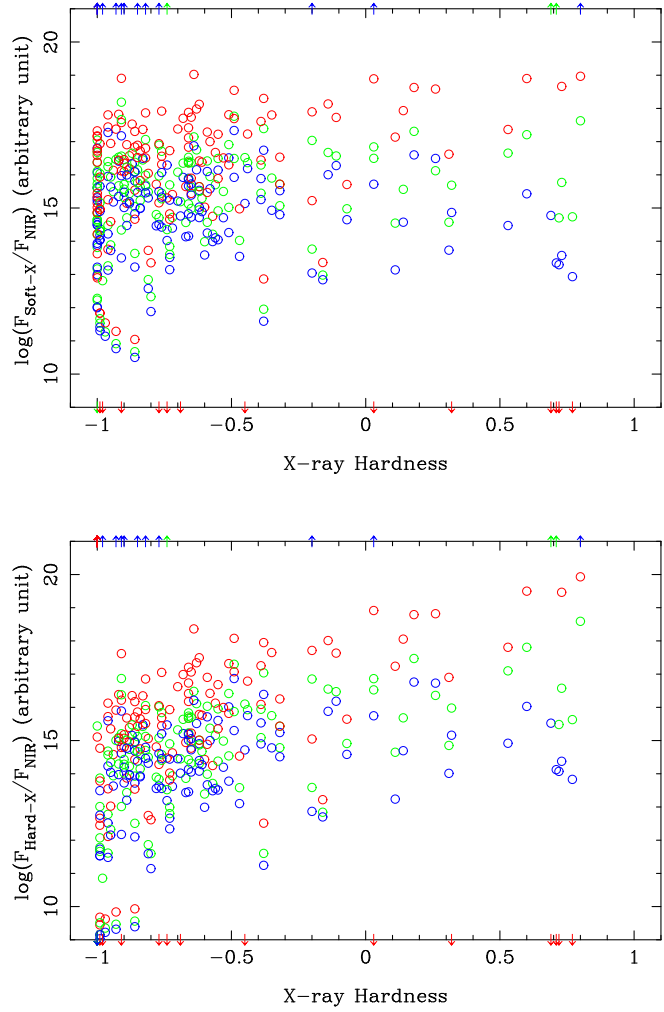}
   \end{center}
\caption{X-ray spectral hardness versus the  X-ray and NIR flux ratio.
Top panel gives the soft X-ray flux over NIR flux ratio,  and the bottom is for the hard X-ray flux.
Red, green and blue colors indicate $J$, $H$ and $K_S$ band fluxes, respectively.
}
\label{HR_Fx_FIR}
\end{figure}

To better understand   properties of the Galactic point sources with NIR counterpart,
we have made a NIR color-color diagram (Fig.\  \ref{color_color}).
Also, we  investigate  correlation between the X-ray fluxes 
and the NIR magnitudes (Fig.\  \ref{X-ray_NIR_corr}), and correlation between the X-ray spectral hardness
and the X-ray to NIR flux ratios (Fig. \ref{HR_Fx_FIR}).
From Fig.\    \ref{color_color} to \ref{HR_Fx_FIR}, we notice the following 
X-ray and NIR characteristics of these sources:
\begin{enumerate}
\item On the NIR color-color diagram,  soft X-ray sources  are mostly on the main-sequence track (Fig.\ \ref{color_color}),
      while medium and hard sources are more scattered.
\item  Soft X-ray and NIR fluxes are correlated   (Fig. \ref{X-ray_NIR_corr} top), and the correlation is along the slope expected for
       the constant luminosity sources at different distances. 
On the other hand, such a  correlation is not clearly seen between the hard X-ray and NIR fluxes
         (Fig. \ref{X-ray_NIR_corr} bottom).
\item  The X-ray spectral hardness and the {\em hard}\/ X-ray to NIR flux ratio shows a correlation
       (Fig. \ref{HR_Fx_FIR} bottom), while such a correlation is not obvious
       between the X-ray spectral hardness and  the {\em soft}\/  X-ray to NIR flux ratio (Fig. \ref{HR_Fx_FIR} top).
\end{enumerate}

The fact (1), together with the thin thermal X-ray spectra (Fig.\  \ref{Souce_Spectra}), 
confirms stellar origin of the soft X-ray sources.
The above fact (2)  suggests that the observed 
 soft X-ray and NIR flux distributions  are both  explained primarily due to various source distances.

The above fact (3)  suggests that 
cataclysmic variables, hard X-ray sources, can emit hard X-rays more efficiently than
X-ray active stars at a given NIR flux.
Also, the wide distribution of the hard X-ray sources on
the NIR color-color diagram (Fig.\  \ref{color_color})
looks  similar to that of the known cataclysmic variables (Hoard et al.\ 2002;  Cutri et al.\ 2005).

What about origin of the sources without NIR counterpart?
Let's first consider the soft X-ray sources.  Background AGNs
are unlikely, since they are almost completely absorbed in the soft X-ray band (Fig.\ \ref{logN-logS} bottom).
It is suggested that  isolated neutron stars may be dim soft X-ray
sources without optical/NIR counterparts (e.g., Popov et al.\ 2000).  
However, there is hardly an isolated neutron star among our  38 soft X-ray sources
without NIR counterparts, since expected surface number density of such isolated neutron star is 
 one per square degree (Popov et al.\  2000), namely, $\sim0.07$ in our {\it Chandra} field.
Consequently it will be more natural to assume that those soft X-ray sources without NIR counterpart
are X-ray active stars whose NIR fluxes are below our sensitivity, because of the
intrinsic low NIR luminosities and/or large distances.

Regarding the hard X-ray sources without NIR counterpart (55 among the total 69  
hard sources),  most of them are considered to be background AGNs.  In fact,
the Galactic flux attenuation is minor  (Fig.\ \ref{logN-logS} top), thus we {\em must}\/ be 
observing almost all  the background AGNs on the Galactic plane.
In fact, our 2 -- 10 keV $\log N - \log S$ curve is successfully modeled by
assuming that all the hard X-ray sources without NIR counterpart are extragalactic,
and those with NIR counterpart are Galactic
(Section \ref{population}).



\subsection{Modeling the 2-- 10 keV  $\log N - \log S$ Curve  on the Galactic Plane }
\label{population}
We are going to model the observed 2-- 10 keV  $\log N - \log S$ curves  on the Galactic plane
 by {\em Chandra},  {\em XMM-Newton} and {\em ASCA},  
following earlier attempts by, e.g.,  Grimm, Gilfanov \& Sunyaev (2002) and Hands et al.\ (2004).
We assume three populations, high $L_x$ population (neutron star binaries),
low $L_x$ population (white dwarf binaries) and extragalactic sources.
We adopt the same Galactic disk model by Hands et al.\ (2004), namely, the Galactic source
population depends on the Galactic radius $R$ and hight $z$ as,
\begin{equation}
N [{\rm pc}^{-3}] = N_0[{\rm pc}^{-3}] \;  \exp(-R/R_G) \exp(-z/200\; {\rm pc}),
\end{equation}
where $R_G$ is the Galactocentric radius 8.5 kpc, and $N_0$ is determined from the
differential luminosity function, as 
\begin{equation}\label{LfunctionEq}
N_0 = \int_{L_{min}}^{L_{max}} A L^{-\alpha} dL.
\end{equation}
The maximum diameter of the Galactic disk is assumed 20 kpc, and the hydrogen density
of 0.55 cm$^{-3}$ (Hands et al.\ 2004).  We integrate the sources in the line of sight
($l=28.\arcdeg5$) until the edge of the Galactic plane (distance will be 27 kpc).
Note that in the small {\em Chandra}\/ field of view ($\sim17\arcmin$), the vertical
distance from the Galactic plane is $\sim130$ pc even at the edge of the Galaxy, smaller than the 200 pc Galactic scale height.
Therefore, if we assume the constant luminosity sources, the $\log N - \log S$ curve is expected to have
the slope $ -1.5$ only determined by  observing volume increase.
For the extragalactic sources attenuated by Galactic extinction  ($\sim 6 \times 10^{23}$ cm$^{-2}$),
we assumed the following functional shape (Ueda et al.\ 1999; Giacconi et al.\ 2001):
\begin{equation}
N(>S) = \left\{\begin{array}{lll}
               25   \;(S/4.9 \times 10^{-14}) ^{-1.5} \; \; &{\rm for}& S \geq 2.6 \times 10^{-14} \; {\rm erg\; cm^{-2}\; s^{-1}}\\
               1200 \;(S/1.4 \times 10^{-15}) ^{-1.0} \; \; &{\rm for}& S  <   2.6 \times 10^{-14} \; {\rm erg\;  cm^{-2} \; s^{-1}}.\\
               \end{array}
        \right.
\end{equation}

\begin{figure}[ht]
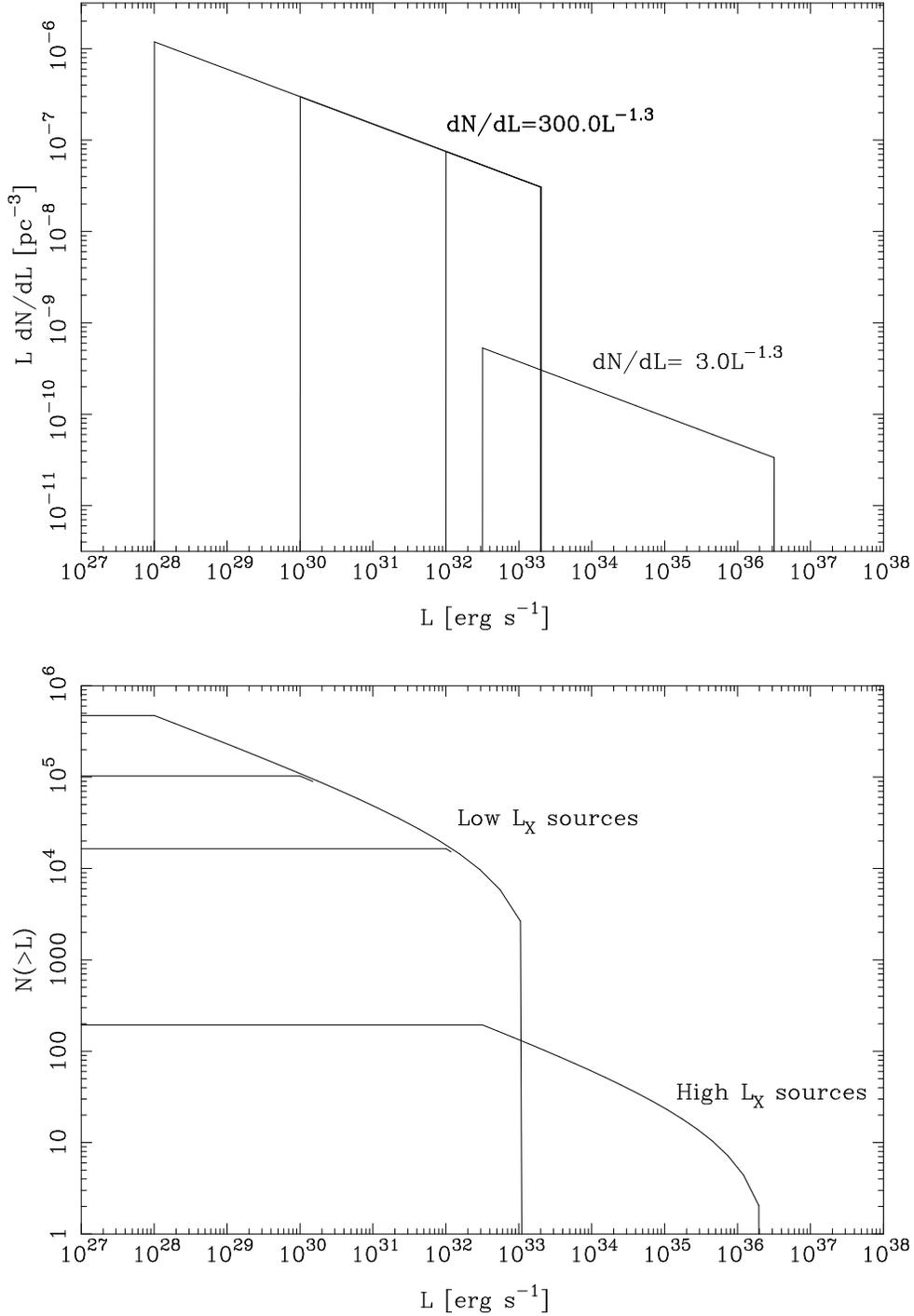

   \begin{center}
     \includegraphics[angle=270,width=13cm]{f23a.eps}
   \end{center}
   \begin{center}
     \includegraphics[angle=270,width=13cm]{f23b.eps}
   \end{center}
\caption{Top: Differential X-ray source luminosity functions
we adopt to model the $\log N-\log S$ curve (see Fig.\ \ref{logNlogSmodel}).
We consider distinct high luminosity  and low luminosity source populations.
For the low luminosity  source population, we try three different lower
cut-off luminosities.  Bottom: 
Cumulative luminosity functions obtained by
integrating over the Galactic source distribution.
}
\label{Lfunction}
\end{figure}

Now we are free to choose four parameters in equation (\ref{LfunctionEq}) for high $L_x$ population and
low $L_x$ population each.  For the  high $L_x$ population, we took $A=3, \alpha=1.3$, $L_{min} =10^{32.5}$
erg s$^{-1}$ and $L_{max}=10^{36.5}$ erg s$^{-1}$ (Fig.\ \ref{Lfunction}). These values are similar to 
those used by 
Grimm, Gilfanov \& Sunyaev (2002) and Hands et al.\ (2004). Integrating the luminosity
function  over the Galactic plane, there will
be $\sim200$ such neutron star binaries, which is reasonable (Fig.\ \ref{Lfunction}).
In any case, our {\em Chandra} observation is not sensitive to the choice of high $L_x$ population
parameters, since they affect only bright sources above $\sim 10^{-12}$ erg s$^{-1}$ cm$^{-2}$.

\begin{figure}[ht]
   \begin{center}
     \includegraphics[width=13cm]{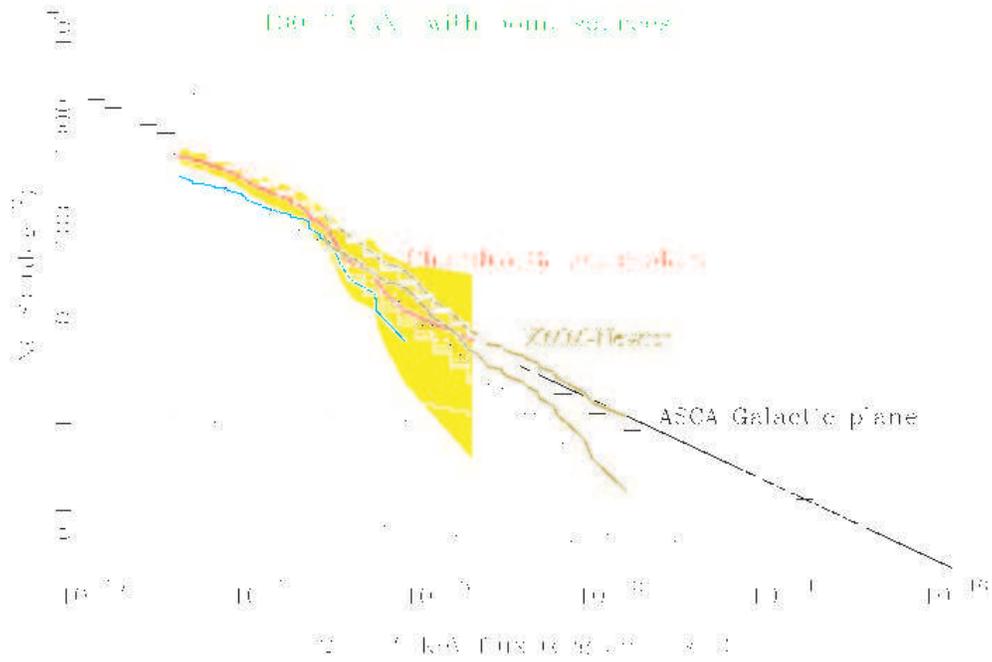}
   \end{center}
\caption{Modeling of 2 -- 10 keV $\log N-\log S$ curves on the
Galactic plane obtained with {\it Chandra} (same as in Fig.\ \ref{logN-logS}), 
{\it XMM} (Hands et al.\ 2004)
and {\it ASCA} (Sugizaki et al.\ 2001).
Galactic luminosity functions  in Fig.\ \protect\ref{Lfunction} are adopted.
As in Fig.\ \ref{logN-logS}, the line in cyan indicates the sources with near-infrared
counterparts.  Note, three different lower cut-off luminosities in the low luminosity
source population (Fig.\ \ref{Lfunction} top) are reflected in only small
differences in the  $\log N - \log S$ curves.
}
\label{logNlogSmodel}
\end{figure}

On the other hand, {\em Chandra} $\log N - \log S$ curve below $2 \times 10^{-13}$ erg s$^{-1}$ cm$^{-2}$ is significantly
dependent on the low $L_x$ population. In particular, we are sensitive to $L_{max}$ and the normalization there.
We found the choice of $A=300, \alpha=1.3$ and $L_{max}=2 \times 10^{33}$ erg s$^{-1}$ (Fig.\ \ref{Lfunction}) can explain the observed
{\em Chandra}, {\em XMM-Newton} and {\em ASCA} 
$\log N - \log S$ curves reasonably well (Fig.\ \ref{logNlogSmodel}).  
In particular, the low $L_x $ population model curve can approximate the {\em Chandra}  $\log N - \log S$ curve of
the hard X-ray sources with NIR counterpart (line in cyan in Fig.\ \ref{logN-logS} and \ref{logNlogSmodel}),
which is reasonable since these sources are considered to be Galactic (see also Section \ref{origin_point_source}).
We tried three very different $L_{min}$=$10^{28}, 10^{28} $ and $10^{30}$ erg s$^{-1}$ (Fig.\ \ref{Lfunction}), but resultant differences in $\log N - \log S$ are  hardly noticeable (Fig.\ \ref{logNlogSmodel}).
If integrated over the Galactic plane, we need at least $10^4$ such dim X-ray sources below $2\times10^{33}$ erg s$^{-1}$ (Fig.\ \ref{Lfunction}).
There may be orders of magnitude more such dim Galactic sources depending on $L_{min}$, but we may not constrain from the
{\em Chandra} $\log N - \log S$ analysis above $\sim3 \times 10^{-15}$ erg s$^{-1}$ cm$^{-2}$.

In summary, from {\em Chandra} and {\em XMM-Newton} $\log N-\log S$ analysis, it is no doubt that
there is a low $L_x$ population of Galactic sources below $\sim 2\times10^{33}$ erg s$^{-1}$ and greater than $10^4$ in number,
in addition to the well-established bright neutron star population.  From our X-ray and NIR study, these 
dim Galactic sources are most likely to be cataclysmic variables (Section \ref{origin_point_source}).
Presence of such dim and numerous cataclysmic variables had been  in fact expected  
(e.g., Mukai \& Shiokawa 1993; Verbant et al.\ 1997; Watson 1999), 
but with {\em Chandra} and {\em XMM-Newton} we are for the first time  
able to  measure their population and X-ray  characteristics accurately.

\subsection{Absolute GRXE Flux and Comparison with Other Measurements}

\label{comparison}

Excluding the point sources in our field of view brighter than $\sim 3 \times 10^{-15}$ erg s$^{-1}$ cm$^{-2}$
(2 -- 10  keV) or  $\sim 2 \times 10^{-16}$ erg s$^{-1}$ cm$^{-2}$ (0.5 -- 2 keV), 
we have extracted and studied the Galactic diffuse emission spectra. Using the best-fit spectral 
model
(Section \ref{diffuse_study}), we determine the observed
diffuse emission flux as  $6.5 \times 10^{-11}$ erg s$^{-1}$ cm$^{-2}$ deg$^{-2}$
(2 -- 10 keV) and  $8.7 \times 10^{-12}$ erg s$^{-1}$ cm$^{-2}$ deg$^{-2}$ (0.5 -- 2 keV).
The soft X-ray flux we measured might  be affected by uncertainty of the background subtraction (Section \ref{seperation}).
On the other hand, our measurement in the hard X-ray band is  considered to be 
most precise to date with  accurate background subtraction,  well-determined mirror response 
and, most of all,   hardly contamination from point sources.

Let's compare our measurement in the 2 -- 10 keV band with those by {\em ASCA} or {\em XMM-Newton}.
In Table \ref{compare_xmm_asca}, we show 
the 2 -- 10 keV GRXE fluxes  measured by {\em ASCA} (Kaneda et al.\ 1997; Sugizaki et al.\ 2001),  {\em XMM-Newton} 
(Hands et al.\ 2004) and  {\em Chandra} (present paper).
We should be careful that point source sensitivities  in these measurements are very different, 
and the GRXE fluxes thus obtained necessarily contain contributions from both diffuse emission and 
point sources
below the thresholds.  If  all the resolved point sources are included,
our {\em Chandra} observation gives the GRXE flux  $7.4 \times 10^{-11}$ erg s$^{-1}$ cm$^{-2}$ deg$^{-2}$
in 2 -- 10 keV, where the brightest point source  (Source 200 in Table 1) has the flux
$\sim 2 \times 10^{-13}$ erg s$^{-1}$ cm$^{-2}$. In the {\em XMM-Newton}  survey (Hands et al.\ 2004) where
point sources between $2\times10^{-14}$ to $2\times10^{-12} $ erg s$^{-1}$ cm$^{-2}$ deg$^{-2}$
are detected, if we exclude the point sources above $\sim 2 \times 10^{-13}$ erg s$^{-1}$ cm$^{-2}$ (brightest 
{\em Chandra} source flux), the GRXE flux is  $9.3 \times 10^{-11}$ erg s$^{-1}$ cm$^{-2}$ deg$^{-2}$.
The {\em XMM-Newton} GRXE flux is $\sim25 \%$ higher than our {\em Chandra} flux, but considering 
the lower Galactic longitudes of the {\em XMM-Newton} survey, which should result in  higher GRXE fluxes,
we consider the {\em Chandra} and {\em XMM-Newton} flux agreement pretty well.

On the other hand, the two independent {\em ASCA} surveys (Kaneda et al.\ 1997; Sugizaki et al.\ 2001) report  much
smaller GRXE fluxes than {\em Chandra} and {\em XMM-Newton} (Table \ref{compare_xmm_asca}).
We suspect this might be due to systematic effects of the {\em ASCA} X-ray telescope;  when largely extended
sources are observed with  {\em ASCA}, stray-lights from outside of the field of view ($\sim25$ arcmin radius)
are significant.  This effect is taken into account in both Kaneda et al.\ (1997) and  Sugizaki et al.\ (2001) to
calculate  the GRXE flux, such that the diffuse flux {\em per sky area}\/ is derived {\em assuming}\/  the diffuse emission is uniformly distributed over 
the $1.\arcdeg5 $ radius. In reality, the GRXE scale height may not be as large as $1.\arcdeg5 $, in that case
{\em ASCA} measurements {\em underestimate}\/ the GRXE surface flux per sky area.

\subsection{Ultimate Point Source Contribution to GRXE}
\label{PointSourceContribution}
Let's consider if the Galactic ``diffuse'' emission we obtained by removing resolved {\em Chandra} point sources
may be accounted for by superposition of still 
dimmer point sources below our detection limits.  
The total energy flux observed in our {\em Chandra}\/ field,  $F_{total}$, may be expressed as follows:
\begin{equation}\label{eq1}
F_{total} [{\rm erg\; s^{-1}\; cm^{-2}\; deg^{-2}}] = F_{S \leq S_{th}} + \int_{S'>S_{th}}^{S_{max}}S'\;\frac{dN(>S')}{dS'} dS',
\end{equation}
where the first term in the right hand side includes
the diffuse emission and the contribution of the point sources equal to or dimmer than the threshold flux $S_{th}$.
The second term is the contribution from the point sources brighter than  $S_{th}$.
We may define the hypothetical  point source number density,
\begin{equation}\label{eq2}
N(S_{th}) \equiv F_{S \leq S_{th}} /S_{th}, 
\end{equation}
whose meaning is as follows: if point source detection above flux $S_{th}$ is complete,
in addition to the  sources brighter than $S_{th}$, the hypothetical $N(S_{th})$ sources having  the same flux $S_{th}$
would account for the total X-ray flux observed in the field of view.
In the present case,  we have set the threshold flux $\sim 3 \times 10^{-15}$ erg s$^{-1}$ cm$^{-2}$
(2 -- 10  keV) and  $\sim 2 \times 10^{-16}$ erg s$^{-1}$ cm$^{-2}$ (0.5 -- 2 keV), then
the first term in equation (\ref{eq1}) is the ``diffuse'' flux we determined earlier.
 The second term in  Eq. \ref{eq1} can be obtained by integrating the
detected point source fluxes above $S_{th}$; for the same  threshold fluxes,
they are $\sim 8.5 \times 10^{-12}$ erg s$^{-1}$ cm$^{-2}$ deg$^{-2}$ (2 -- 10 keV)
and $\sim 1.0 \times 10^{-12}$ erg s$^{-1}$ cm$^{-2}$ deg$^{-2}$ (0.5 -- 2 keV), respectively.
Therefore, the total observed flux in the field view is 
$\sim 7.4  \times 10^{-11}$ erg s$^{-1}$ cm$^{-2}$ deg$^{-2}$ (2 -- 10 keV)
and $\sim 9.7 \times 10^{-12}$ erg s$^{-1}$ cm$^{-2}$ deg$^{-2}$ (0.5 -- 2 keV), respectively.

In Fig.\  \ref{logN-logS} and \ref{logNlogSmodel}, 
we show  $N(S_{th})$ defined in Eq.\ \ref{eq2} for different values of $S_{th}$. 
In order to account for the total  X-ray fluxes 
in the field of view with 100 \% of point sources,  it is required that the 
$\log N - \log S$ curves rapidly steepen by an order of magnitude somewhere below our sensitivity limits,
which is extremely unlikely.
In particular, even if we significantly extrapolate the low $L_x$  luminosity function to
the dimmer side (Fig.\ \ref{Lfunction}), increase of  $\log N - \log S$ curve is very tiny (Fig.\ \ref{logNlogSmodel}).

\begin{figure}[ht]
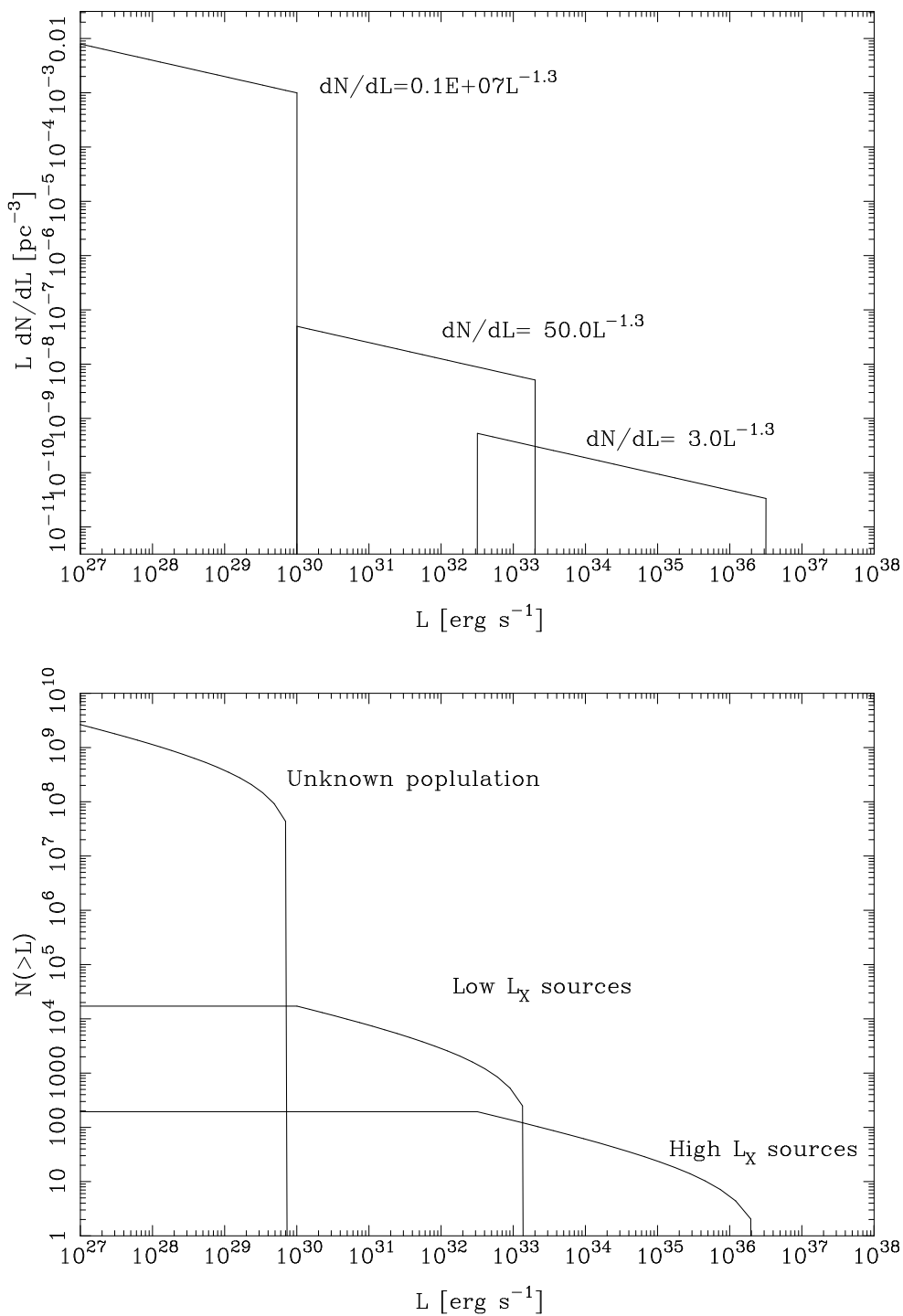

   \begin{center}
     \includegraphics[angle=270,width=13cm]{f25a.eps}
   \end{center}
   \begin{center}
     \includegraphics[angle=270,width=13cm]{f25b.eps}
   \end{center}
\caption{Top: Differential X-ray source luminosity functions
assuming a hypothetical population of very dim sources ($L_X\leq10^{30}$
erg s$^{-1}$) to model the $\log N-\log S$ curve (see Fig.\ \ref{logNlogSmodel_trial}).
Bottom: Cumulative luminosity functions obtained by
integrating over the source distributions above.
}
\label{Lfunction_trial}
\end{figure}

\begin{figure}[ht]
   \begin{center}
     \includegraphics[angle=270,width=13cm]{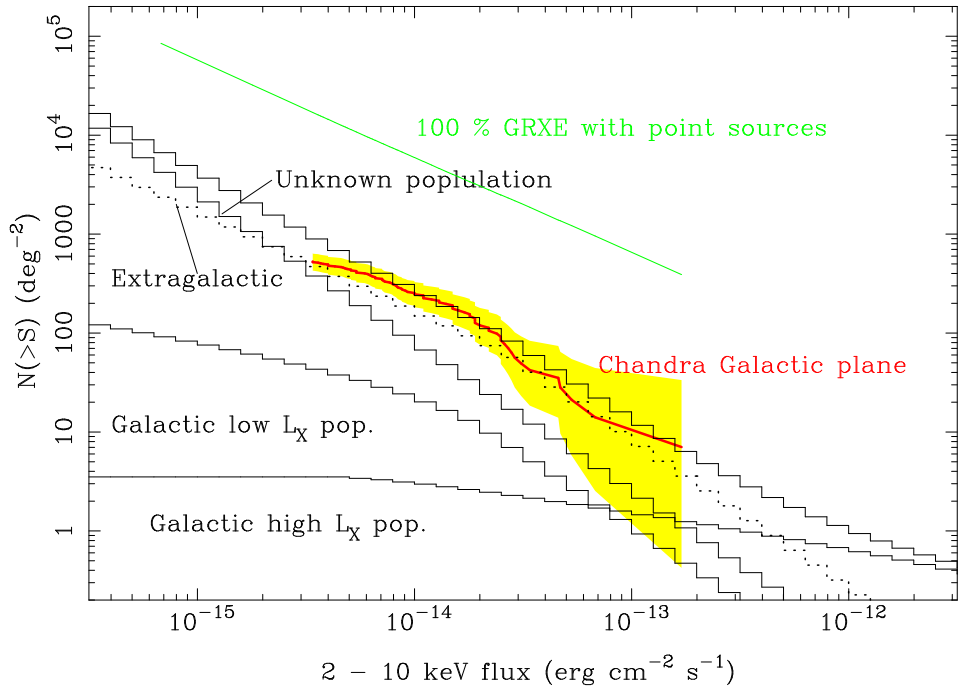}
   \end{center}
\caption{A trial modeling of 2 -- 10 keV $\log N-\log S$ curve on the
Galactic plane obtained with {\it Chandra} (same as in Fig.\ \ref{logN-logS}), 
assuming the hypothetical  luminosity functions in Fig.\ \ref{Lfunction_trial}.
Note, introducing extremely dim and numerous point sources below $L_X\leq10^{30}$
erg s$^{-1}$ hardly contributes to the  Galactic Ride X-ray Emission
flux.}
\label{logNlogSmodel_trial}
\end{figure}

Furthermore, we examine if introducing another unknown, still dimmer Galactic source population
might explain the 100 \% of  the GRXE flux in 2 -- 10 keV. Now we consider luminosity functions as shown in Fig.\ \ref{Lfunction_trial}.
There is a hypothetical source population below $10^{30}$ erg s$^{-1}$, and the total number of 
such sources in the Galaxy is $\gtrsim10^9$, in addition to neutron star (high $L_x$) and white dwarf (low $L_x$) populations.
Although combination of these  populations can more or less explain the observed {\em Chandra} $\log N - \log S$ curve 
above $3 \times 10^{-15}$ erg s$^{-1}$ cm$^{-2}$ (Fig.\ \ref{logNlogSmodel_trial}),
it is not sufficient at all to explain the 100 \% of GRXE.  The point is that any $\log N - \log S$ curves 
of Galactic source populations, whatever the luminosity function is,  cannot have a slope steeper than $-1.5$.
Hence, no Galactic source population can explain both the observed {\em Chandra}
$\log N -  \log S$ curve 
and the 100 \% of the GRXE flux  sinultaneously. Consequently, we conclude GRXE is primarily diffuse emission.

\subsection{Origin of the Galactic Diffuse  X-ray Emission}\label{origin_diffuse} 
We found that   GRXE has a truly diffuse origin,  then the question
is how to produce and maintain such high energetic plasma.  
There are obvious problems in interpreting GRXE in terms of simple equilibrium
thermal plasma, such that the plasma temperature needed to explain the observed spectra,  $kT \approx$ 5 -- 10 keV, 
is much higher than can be bound by Galactic gravity (Warwick et al.\ 1985).
Also,  the energy density of GRXE, $\sim$10 eV/cm$^3$, 
is one or two orders of magnitude 
higher than those of other constituents in the interstellar space, 
such as cosmic rays, Galactic magnetic fields, or ordinary
interstellar medium  (Koyama et al.\ 1986; Kaneda et al.\ 1997).
Currently there are no accepted theoretical models that can explain the origin
of  GRXE.  Some argue 
that the interstellar magnetic field is playing a significant role to heat and
confine the hot plasma (Tanuma et al.\ 1999).
Others propose  that the interstellar medium is mainly responsible for GRXE and gamma-ray emission, 
via  interactions with, for instance,  low energy cosmic-ray electrons 
(Valinia et al.\ 2000b), in situ accelerated quasi-thermal electrons (Dogiel et al.\ 2002;  Masai et al.\ 2002),
or heavy  ions (Tanaka, Miyaji, \& Hasinger 1999).  Galactic particle acceleration is considered to be taking place 
in supernova remnants.  
In fact, serendipitous discovery of the hard X-ray
emitting supernova remnant  AXJ 1843.8--0352 in our field (Bamba et al.\ 2001; Ueno et al.\ 2003) 
strongly suggests a close tie between GRXE and supernova remnants.

Various theoretical models of GRXE have to be tested through observations.
Different heating or acceleration mechanism of the plasma will result in different
plasma conditions, which are reflected in the emission lines.
Therefore, from precise measurements of the GRXE emission lines,
we may in principle diagnose the plasma conditions and constrain the theoretical models.
In particular, iron line spectroscopy is essential.  We have shown that the GRXE iron line
central energy is $6.52\pm^{0.08}_{0.14}$ keV (Section \ref{diffuse_study}), significantly lower than what expected
from thermally equilibrium plasma ($\sim$6.67 keV).  Although from our data
we could not distinguish if this iron line is really a 
single line or composite of two or three lines, the  {\it Chandra} Galactic center
diffuse spectrum  clearly indicates the three emission lines from low ionized iron, He-like
iron and H-like iron (Muno et al.\ 2004).  If we assume similar origins of Galactic center
and plane diffuse emission (see Section \ref{GCcomparison}),  then the three line interpretation of GRXE iron line emission
seems plausible (see also Tanaka 2002).  If this is the case, the 6.4 keV line is considered from
fluorescence in cool interstellar medium, which may be induced, for instance, by
low energy cosmic-ray electrons (Valinia et al.\ 2000b).  The He-like and H-like lines
may be  from hot thermal equilibrium plasma (Valinia et al.\ 2000b),   charge exchange
process of iron ions (Tanaka, Miyaji, \& Hasinger 1999; Tanaka 2002),  or  recombination cascades of quasi-thermal electrons (Masai et al.\ 2002).
In  the second case,  the lines are expected to be significantly broadened by the iron nucleus bulk motion (Tanaka et al.\  2000).
%
In the last case, the emission lines are accompanied by recombination continuum above He-like and H-like iron K-edges (Masai et al.\ 2002).
These three cases may be distinguished from precise spectral observation in the iron line/edge energy range.

We emphasize that the precise  iron line diagnostic (including line intrinsic width measurement) is a key to resolve 
origin  of the Galactic center and Galactic plane diffuse emission.  In this context, planned Galactic center
observations with {\em Astro-E2}\/ XRS, the first X-ray microcalorimeter in space with $\sim6$ eV resolution, 
will be an enormous help. GRXE may be too dim for {\em Astro-E2}\/ XRS (our simulation suggests that
a million second exposure is required), but 
we believe that  the long standing mystery of GRXE will be certainly
solved by future calorimeter observations with much higher throughputs and better
spectral resolution, expected to be made by {\em Con-X}, {\em NEXT} and/or{\em XEUS}.

\subsection{Comparison with the Galactic Center}
\label{GCcomparison}
It will be interesting to compare our point source populations in the Galactic plane ($l \approx 28.\arcdeg5$) with
those at   Sgr A ($l \approx 0\arcdeg$) and Sgr B2 ($l \approx 0\arcdeg.5$).
We have analyzed the {\it Chandra} Galactic center Sgr B2 region data  (observed in 2000 March 29 for 100 ksec,
obsID=944; Murakami, Koyama and Maeda 2001) in a similar manner to our Galactic plane data analysis,
and made $\log N - \log S$ curves for both energy bands in  Fig.\ \ref{logN-logS}.  The $\log N - \log S$ curves in the Sgr A region
are  also plotted (from Muno et al.\  2003).
In the hard energy band, the source number density increases dramatically toward the Galactic center, 
significantly exceeding that on the Galactic plane and extragalactic one.  This indicates that there
are much more Galactic hard point sources in the Galactic center region than
in the Galactic plane.  
In the soft band, on the other hand, the source number density at the lowest flux
level is not so different on the Galactic plane and at the Galactic center regions.
This is probably because the faintest observable soft sources are mostly located 
in our neighborhood, so that the direction toward the Galactic center or Galactic plane 
will not make a big difference.  In fact, such dim soft sources at the Galactic
 center will be more significantly absorbed ($N_H \approx 10^{23}$ cm$^{-2}$) than
those  extragalactic sources in our Galactic  plane field ($N_H \approx 6 \times 10^{22}$ cm$^{-2}$) , and  thus hardly detected.
 

We also compare  diffuse emission from the Galactic plane and Galactic center.
Considering the diffuse spectral similarity 
on the Galactic plane (Section \ref{diffuse_study}) and Galactic center (Muno et al.\ 2004),
we are tempted to conclude  that they have similar origins (see also Tanaka 2002).  Recent INTEGRAL  
observations have detected  hard X-ray emission above  20 keV in the Galactic center region
whose centroid is slightly offset of Sgr A$^*$  (B\'elanger et al.\ 2004).
This suggests
that a non-thermal hard-tail of  the Galactic center diffuse spectrum is extended
above 20 keV.
GRXE also has a power-law hard-tail component which extends above $\sim$20 keV
(Yamasaki et al.\ 1997; Valinia and Marshall 1998).
On the other hand, strong diffuse  gamma-ray ($\sim$ 100 keV -- 1 MeV) emission
is observed from the Galactic center and plane region 
(e.g., Gehrels and Tueller 1993; Skibo et al.\ 1997; Valinia et al.\ 2000a;
Strong et al.\ 2003), which is suggested to have a non-thermal origin.
Intriguingly,  the Galactic center and plane diffuse  hard X-ray  components
seem to be  smoothly connected to the gamma-ray components.
We suspect there is a common  physical mechanism in the Galactic center and plane diffuse
emission to produce hard X-ray and gamma-ray emission from several keV to $\sim$ MeV.

\section{Conclusion}

Using {\it Chandra} ACIS-I, we have carried out a deep X-ray observation (0.5 -- 10 keV) on a typical Galactic
region  at $(l,b) \approx (28.^\circ5,  0.^\circ0)$ within  $\sim $250 arcmin$^2$ to study
characteristics of Galactic Ridge X-ray Emission (GRXE),
followed by a NIR identification observation with NTT/SOFI at ESO.  Our main results are summarized below:
\begin{enumerate}
\item We have detected 274 new X-ray sources   (4 $\sigma$ confidence) down to 
 $\sim3 \times 10^{-15} $  erg s$^{-1}$ cm$^{-2}$ in 2 -- 10 keV or   $\sim2 \times 10^{-16} $ erg s$^{-1}$ cm$^{-2}$ in 0.5 -- 2 keV.
Only 26 sources are detected both in the soft and hard bands.  In the SOFI field, 
\mbox{83 \%} of the soft sources
are identified in NIR, while only \mbox{22 \%} of the hard sources have NIR conterparts.  Most of the soft X-ray sources
are considered to be X-ray active stars, while most of the unidentified hard X-ray sources are extragalactic. 

\item Only $\sim$ \mbox{10 \%} of the observed  X-ray flux in the {\it Chandra} field is accounted for by the sum of point source fluxes.
      Even if we assume an unknown population of much dimmer and numerous Galactic sources, the observed GRXE  flux is not explained.
      Therefore,  we conclude that GRXE  has  truly diffuse origin, confirming our
      early report using half of the current {\em Chandra} data (Ebisawa et al.\ 2001) and in agreement with the {\em XMM-Newton}\/ Galactic survey
      with a larger sky coverage (Hands et al.\ 2004).

\item Soft X-ray sources exhibit thin thermal spectra, characteristics of active stars. 
      In fact, they follow  the track of main-sequence stars on the NIR color-color diagram.
      Small number of the hard X-ray sources with NIR  counterpart 
      exhibit a narrow iron emission line at 6.67 keV as a signature of the  Galactic cataclysmic variables.
      To explain the observed 2 -- 10 keV $\log N - \log S $ curve, we suppose there are at least $10^4$ cataclysmic variables
      in the Galactic plane  dimmer than  $\sim 2\times10^{33}$ erg s$^{-1}$.

\item Removing contamination of point X-ray sources brighter than $3 \times 10^{-15}$ erg s$^{-1}$ cm$^{-2}$ (2 -- 10 keV), 
      we have precisely measured the Galactic diffuse X-ray emission flux  as $6.5 \times 10^{-11}$  erg cm$^{-2}$ s$^{-1}$ deg$^{-2}$ in 2 -- 10 keV.
      The  energy spectrum of the diffuse emission can be  modeled with a two temperature non-equilibrium ionization model, such that the soft,
      less absorbed component is more highly ionized than the hard, more absorbed component.       

\item We have measured the diffuse iron emission line energy as $6.52\pm^{0.08}_{0.14}$ keV 
(\mbox{90 \%} error).  This is significantly
      lower than what is expected from thermally equilibrium plasma (6.67 keV).  This shift of the iron line energy may be 
      explained  either by  non-equilibrium ionization of the plasma, or hybrid of the 6.4 keV fluorescent line and the
      6.67 keV line from equilibrium plasma.
\end{enumerate}

\acknowledgments

We are grateful to R. Mushotzky, Y. Yang and K. D. Kuntz for useful comments on the
{\it Chandra} data analysis and discussion. We thank the second referee for his/her insight of the
ACIS intrinsic background spectral feature.
Support for this work was provided by the National Aeronautics and Space Administration (NASA) through 
{\em Chandra}\/ Guest Observer program  (Sequence number 900021 and 900125)
issued by the Chandra X-ray Observatory Center, 
which is operated by the Smithsonian Astrophysical Observatory for and on behalf of NASA  under contract NAS8-03060.
This publication is based on observations 
collected at the European Southern Observatory, Chile (ESO N$^\circ$ 69.D-0664).
Also, this paper makes use of data products from the Two Micron All Sky Survey, 
which is a joint project of the University of Massachusetts and the Infrared Processing 
and Analysis Center/California Institute of Technology, funded by NASA and the National Science Foundation.
An image used in this paper was  produced with Montage, an image mosaic service
supported by the NASA Earth Sciences Technology Office Computing
Technologies program, under Cooperative Agreement Notice NCC 5-6261
between NASA and the California Institute of Technology.
M.T. is financially supported by the Japan Society for the Promotion of Science.


\clearpage

\clearpage

\clearpage

\clearpage

\clearpage

\clearpage

\clearpage

\clearpage

\clearpage

\clearpage

\clearpage

\clearpage

\clearpage

\clearpage

\clearpage

\clearpage

\clearpage

\clearpage

\clearpage

\clearpage

\clearpage

\clearpage

\clearpage

\clearpage

\clearpage
\pagestyle{empty}
\thispagestyle{empty}
\begin{deluxetable}{lccccccccccccc}
\tabletypesize{\tiny}
\rotate
\tablewidth{0pt}
\tablecolumns{16}
\tablecaption{Detected Source List}
\tablehead{\colhead{ID\tablenotemark{a}}&\colhead{AO}&\colhead{R.A.}&\colhead{Dec.}&\colhead{Error}&\colhead{Significance}&\colhead{Normalized counts}&\colhead{Hardness}&\colhead{Flux}&\multicolumn{5}{c}{NIR conterpart}\\\cline{10-14}
\colhead{}&\colhead{}&(hh:mm:ss.ss)&(d:mm:ss.s)&(arcsec)&($\sigma$)&\colhead{(cnts/100ks)}&\colhead{(H-S)/(H+S)}&\colhead{(ergs s$^{-1}$ cm$^{-2}$)}&
\colhead{ID\tablenotemark{b}}&\colhead{seperation}&\colhead{$J$\tablenotemark{c}}&\colhead{$H$\tablenotemark{c}}&\colhead{$K_S$\tablenotemark{c}}\\
\colhead{}&\colhead{}&\multicolumn{2}{c}{(J2000)}&&\colhead{Total/Soft/Hard}&\colhead{Total/Soft/Hard}&&\colhead{0.5-2keV/2-10keV}&&\colhead{(arcsec)}&\colhead{(mag)}&\colhead{(mag)}&\colhead{(mag)}}
\startdata 
1&2&18:42:51.77& $-$3 51 11.2&0.86&5.5/5.1/0.0&27.3/27.1/0.0&-1.00 $\pm$  0.09& 4.0E-16/1.6E-16        &OoF&\nodata & \nodata &\nodata &\nodata \\ 
2$^T$&2&18:42:53.53& $-3$ 49 19.6&0.93&6.9/3.6/6.3&67.7/21.3/48.8&0.39 $\pm$  0.12& 3.5E-16/1.0E-14    &OoF&\nodata & \nodata &\nodata &\nodata\\ 
3&2&18:42:58.30& $-3$ 53 28.1&0.49&11.3/0.0/11.4&79.2/11.0/70.0&0.73$\pm$  0.08& 4.2E-17/3.1E-14       &J18425832-0353271 & 1.00& 17.62  &14.73 & 12.53 \\
4$^T$&2&18:43:00.41& $-3$ 53 48.9&0.36&19.0/18.5/6.0&144.7/112.9/32.1&-0.56$\pm$  0.07& 2.5E-15/1.0E-14&J18430043-0353486 & 0.36& 14.62&  13.65&  12.05 \\ 
5&2&18:43:01.50& $-3$ 50 47.3&0.69&5.0/0.0/5.5&17.8/2.5/15.8&0.73 $\pm$  0.20& 8.8E-18/6.4E-15         &OoF&\nodata & \nodata &\nodata &\nodata\\ 
6&2&18:43:01.83& $-3$ 54 42.4&0.48&7.4/6.7/2.0&45.9/29.8/16.3&-0.29 $\pm$  0.15& 9.5E-16/3.9E-15       &OoF&\nodata & \nodata &\nodata &\nodata\\ 
7&2&18:43:06.14& $-3$ 54 54.8&0.43&4.1/4.6/0.0&20.9/15.2/5.8&-0.45 $\pm$  0.21& 4.3E-16/1.8E-15        &OoF&\nodata & \nodata &\nodata &\nodata\\ 
8&2&18:43:06.88& $-3$ 55 37.0&0.35&8.1/6.0/4.0&30.1/17.0/13.2&-0.12 $\pm$  0.19& 2.1E-16/6.2E-15       &OoF&\nodata & \nodata &\nodata &\nodata\\
\enddata 
\tablenotetext{a}{Sources with  $T$ in the ID column show significant time variations, and those with $E$ are extended.}
\tablenotetext{b}{SOFI ID in Table 1 (starts with S) and/or  2MASS ID (starts with J) of the NIR source within  $1''$ from the Chandra position.  ``No'' means that the source is in the 
                  SOFI field of view, but there are no SOFI counterparts.  ``OoF'' means that the source is out of the SOFI field, and
                  no 2MASS counterparts.}
\tablenotetext{c}{2MASS magnitudes are taken for stars brighter than 10 mag, since SOFI starts to saturate for $<10$ mag (see Fig. \protect\ref{2massSOFIPhoto}).}
\label{srclist_table}
\end{deluxetable} 

\clearpage

\begin{deluxetable}{rllrrrcc}
\tablewidth{0pc}
\tablecaption{SOFI NIR source catalog. }
\tablehead{
\colhead{ID}  & \colhead{R.A.\ (J2000)}  & \colhead{Decl.\ (J2000)}  &\colhead{$J$}  &\colhead{$H$}  &\colhead{$K_S$}  & \colhead{2MASS ID} & \colhead{Chandra ID}\\
\colhead{}    & \colhead{}               & \colhead{}               &\colhead{(mag)}&\colhead{(mag)} &\colhead{(mag)} & \colhead{}         & \colhead{}
}
\startdata   
    1 &18:43:08.380 &$-$03:54:34.12&  -999  & 17.64 &  15.58 &  ...              &    ... \\
    2 &18:43:08.382 &$-$03:52:29.29&  16.69  & 15.33 &  15.09 &  ...              &   ... \\
    3 &18:43:08.382 &$-$03:53:46.51&  17.37  & 16.56 &  16.40 &  ...              &   ... \\
...\\
    8 &18:43:08.451 &$-$03:54:07.29&   -999  & 16.04 &  14.56 &  18430845$-$0354075 &   ...\\
...\\  
   687&18:43:15.299 &$-$03:54:19.16&  10.45  & 10.15 &  10.03 &  18431530$-$0354190 &    24\\
...\\
 1018 &18:43:17.438 &$-$03:57:32.71&  16.54  & 15.25 &  15.13 &  ...              &    30\\
...\\
32397 &18:44:28.474 &$-$04:06:50.97&  15.72  & 13.80 &  12.64 &  ...              &   ...\\
32398 &18:44:28.477 &$-$04:04:33.27&   -999  & -999  &  15.18 &  ...              &   ...\\
\enddata
\tablecomments{
We have detected 32,398 sources from our SOFI fields 
(Fig.\  \protect\ref{SOFI_pointing}).   For each source,
the position, $J$, $H$ and $K_S$ magnitudes, and the 2MASS and/or  {\it Chandra} counterparts are given.
When not detected, $-999$ is put for the magnitude.
The {\it Chandra} source identification number is  defined in Table \ref{srclist_table}.
The complete version of this table is in the electronic
edition of the Journal.  The printed edition contains only a sample.
}
\label{SOFIcatalog}
\end{deluxetable}

\clearpage

\begin{deluxetable}{ccccc}
\tablecolumns{5}
\tablecaption{Summary of the point source characteristics.}
\tablehead{\colhead{}&\colhead{soft}&\colhead{Medium}&\colhead{Hard}&\colhead{Total}\\
\colhead{hardness ratio}&\colhead{$\leq-0.60$}&\colhead{$-0.59$ to $ 0.1$}&\colhead{$\geq0.11$}}
\startdata
All               & 137  &68      &69    & 274\\
Extended\tablenotemark{a}          &    1 & 3      & 0    &4\\
Point Source      & 136  &65      &69    &270\\
\\
Time variation\tablenotemark{b}
                  &    4 &10      &3     &17\\
                  &(3 \%)&(15 \%) &(4 \%)&(6 \%)\\
\\
2MASS counterpart\tablenotemark{b}
                  &  62  &17      &8     &  87 \\
                  &(45 \%)&(26 \%)&(12 \%)&(32 \%)\\
\\
Covered by SOFI\tablenotemark{b}
                  & 106  &47      &49    &202\\
                  &(78 \%)&(72 \%)&(71 \%)&(75 \%)\\
SOFI counterpart\tablenotemark{c}
                  &88    &25      &11     &124\\
                  &(83 \%)&(53 \%)&(22 \%)&(61 \%)\\
\\
NIR counterpart\tablenotemark{d}
                  &98    &30     &14     &142\\
                  &(72 \%)&(46 \%)&(20 \%)&(53 \%)\\\hline
\enddata
\tablenotetext{a}{Sources associated with the thermal blob-like feature CXOU J184357-035441 (Ueno et al.\ 2003).}
\tablenotetext{b}{The percentage gives the fraction among all the point sources, excluding the extended sources.
                  Note that the 2MASS covers the entire {\it Chandra} field, but SOFI does not.}
\tablenotetext{c}{The percentage gives the fraction among the point sources covered by SOFI.}
\tablenotetext{d}{Presence of either 2MASS or SOFI counterparts.
                  The percentage gives the fraction among all the point sources.}
\label{point_source_table}
\end{deluxetable}

\clearpage

\begin{center}
\begin{deluxetable}{cccc}
\tablecaption{Spectral fitting results of the average  spectra with absorbed power-law model}
\tablehead{\colhead{}                 &Soft                 &Medium              &Hard\\
\colhead{hardness ratio}              &$\leq-0.60$          &$-0.59$ to $ 0.1$   &$\geq0.11$}
\startdata
{\em Sources with NIR counterparts}\\
$N_H$ ($10^{22}$ cm$^{-2}$)           &$0.50\pm0.05$        &$0.9\pm0.1$         &$3.5\pm0.4$\\
photon-index                          &$2.8\pm^{0.2}_{0.1}$ &$2.0\pm0.2$         &$0.90^d$\\
$N^a$                                 &$1.5\pm0.2$          &$1.6\pm^{0.4}_{0.3}$&$1.34\pm0.11$\\
reduced $\chi^2$ (dof)                &1.7 (181)            &0.99 (137)          &1.2 (62)\\
observed  flux$^b$ (0.5 -- 10 keV)    &2.1                  &4.7                 &17\\
intrinsic  flux$^c$ (0.5 -- 10 keV) &4.8                  &7.8                 &23\\
\\
\enddata
\tablenotetext{a}{Power-law normalization in $10^{-6}$ photons s$^{-1}$ keV$^{-1}$ cm$^{-2}$ at 1 keV.}
\tablenotetext{b}{Average source flux in units of $10^{-15}$ erg s$^{-1}$ cm$^{-2}$. }
\tablenotetext{c}{Intrinsic source flux in  units of $10^{-15}$ erg s$^{-1}$ cm$^{-2}$ when hydrogen column density is set to null.  }
\tablenotetext{d}{Fixed to the best-fit value obtained for the other group  in the same spectral hardness with/without NIR counterpart.  }
\tablecomments{Errors correspond to 90 \% confidence.}
\label{power_law_fitting}
\end{deluxetable}
\end{center}

\clearpage

\begin{center}
\begin{deluxetable}{ccc}
\tabletypesize{\footnotesize}
\tablecaption{Spectral fitting results of the average  spectra with physical models}
\tablehead{\colhead{}         & With NIR  & Without NIR \\
           \colhead{}         & counterparts  & counterparts}
\startdata
{\em Soft spectra} (two MEKAL model)            &      &  \\
Overall normalization\tablenotemark{a}          & 1.0  &0.52\\
$N_H$ ($10^{22}$ cm$^{-2}$)    &\multicolumn{2}{c}{$1.19\pm0.15$}   \\
$kT$ (keV; soft)               &\multicolumn{2}{c}{$0.25\pm0.02$}   \\
$N$\tablenotemark{b} (soft)                  &\multicolumn{2}{c}{$56$}   \\
$kT$ (keV; hard)               &\multicolumn{2}{c}{$2.2\pm0.5$}   \\
$N$\tablenotemark{b} (hard)                  &\multicolumn{2}{c}{$2.8$ }   \\
reduced $\chi^2$ (dof)         & 1.35 (179)   & 1.85 (30)\\
\enddata
\tablenotetext{a}{The same two temperature MEKAL model parameters are  used for the soft spectra with or without
                  the NIR counterparts with the exception of the overall normalization.
                  Hydrogen density of the plasma is fixed to $n_H = 1$ cm$^{-3}$.  Parameter errors 
                  are for the spectrum with NIR counterparts.}
\tablenotetext{b}{MEKAL model normalization, $10^{-20}/4\pi d^2 \int n_e n_H dV$, where $d$ is
                  distance (cm), $n_e$ and $n_H$ are electron and hydrogen densities (in cm$^{-3}$),
                  respectively.}
\tablenotetext{c}{Power-law normalization in $10^{-6}$ photons s$^{-1}$ keV$^{-1}$ cm$^{-2}$ at 1 keV.}
\tablecomments{
Errors correspond to 90 \% confidence.
The spectral model for the medium  sources without NIR counterpart 
is given in Table \protect\ref{power_law_fitting}.
}
\label{fitting}
\end{deluxetable}
\end{center}
\clearpage

\begin{center}
\begin{deluxetable}{ccccc}
\tabletypesize{\footnotesize}
\tablecaption{Iron line parameters of the Galactic diffuse emission}
\tablehead{                  &\colhead{Energy}& \colhead{Line flux}              & Equivalent width & \colhead{Identification} \\
                             &\colhead{(keV)} & \colhead{(photons s$^{-1}$ cm$^{-2}$ str$^{-1}$)} & \colhead{(eV)}   & \colhead{}}
\startdata
{\em Single line model}$^a$      &                   &                           &                  &              \\
                             &$6.52\pm^{0.08}_{0.14}$&$0.55\pm0.39$              &$170\pm120$       & NEI plasma?\\

{\em Three  line model}$^b$  &                &                                  &                  &              \\
                             &6.40 (fixed)    & $0.6\pm^{0.3}_{0.6}$             &$100\pm^{50}_{100}$     & Fluorescence from low-ionized iron\\
                             &6.67 (fixed)    & $0.4\pm^{0.8}_{0.3}$             &$180\pm^{360}_{140}$     & He-like iron\\
                             &6.97 (fixed)    & $0.5\pm^{0.8}_{0.5}$             &$160\pm^{260}_{160}$     & H-like iron\\
\enddata
\tablenotetext{a}{Reduced $\chi^2$ = 0.83 with dof=4.}
\tablenotetext{b}{Reduced $\chi^2$ = 1.19 with dof=3.}
\tablecomments{
Errors correspond to 90 \% confidence.
Intrinsic line width is fixed to zero, assuming that the lines
are narrower than the instrumental width ($\sim 100$ eV).}
\label{iron_line_fitting}
\end{deluxetable}
\end{center}
\clearpage

\begin{center}
\begin{deluxetable}{cccl}
\tablecaption{Low energy line parameters of the Galactic diffuse emission}
\tablehead{                  \colhead{Energy}& \colhead{Line flux}              & Equivalent width & \colhead{Identification} \\
                             \colhead{(keV)} & \colhead{(photons/s/cm$^2$/str)} & \colhead{(eV)}   & \colhead{}}
\startdata
                $ 1.021\pm^{ 0.009}_{ 0.002}$&  $0.60\pm^{0.14}_{ 0.11}$        &$64\pm^{ 15}_{ 12}$ & \ion{Ne}{10},Ly$\alpha$ (1.02 keV)\\
                $  1.13\pm^{ 0.07}_{ 0.03}$  &  $0.12\pm^{0.12}_{ 0.09}$        &$ 13\pm^{ 13}_{10}$ & Fe L?\\
                $1.32\pm^{ 0.02}_{ 0.01}$    &  $0.32\pm0.08$                   &$ 36\pm10$          & \ion{Mg}{11}, K$\alpha$ (1.35 keV)\\
                $ 1.46\pm^{ 0.04}_{ 0.03}$   &  $0.11\pm^{0.11}_{ 0.08}$        &$ 12\pm^{ 12}_{  9}$& \ion{Mg}{12}, Ly$\alpha$ (1.47 keV)\\
                $1.74\pm^{ 0.06}_{ 0.04}$    &  $<0.32$                         &$<25$               & Low ionized Si?\\
                $1.849\pm^{ 0.003}_{ 0.017}$ &  $0.87\pm^{0.17}_{ 0.46}$        &$110\pm^{ 20}_{ 10}$& \ion{Si}{13}, K$\alpha$ (1.86 keV)\\
                  2.00 (fixed)               &  $0.05\pm^{0.11}_{ 0.05}$        &$6\pm^{ 13}_{ 6}$   & \ion{Si}{14}, Ly$\alpha$  (2.00 keV)\\
                $2.188\pm^{ 0.012}_{ 0.034}$ &  $0.48\pm0.24$                   &$60\pm30$           & Instrumental Au M-complex\\
\enddata
\tablecomments{ Reduced $\chi^2$  = 1.03 with dof = 27.
Errors correspond to 90 \% confidence.
Intrinsic line width is fixed to zero, assuming it is narrower than the instrumental width ($\sim 100$ eV).
}
\label{soft_line_fitting}
\end{deluxetable}
\end{center}

\clearpage

\begin{center}
\begin{deluxetable}{ccc}
\tablecaption{Two NEI model fitting of the Galactic diffuse emission}
\tablehead{                &  \colhead{Fixed abundance model}& \colhead{Free abundance model}}
\startdata
{\em Soft Component}~~~~~~~~~~~~~~~~~~~~\\
                $kT$ (keV)                     &  0.86                 &  $0.59\pm0.09$ \\
                $\log n_et$ (cm$^{-3}$ s)      &  11.1                 &  $11.8\pm0.5$ \\
                Abundance (except Ne, Mg, Si)  &  0.11                 &  $0.044\pm0.037$\\
                Ne abundance                   &  0.11              &$0.30\pm0.2$  \\
                Mg abundance                   &  0.11              &$0.14\pm0.06$  \\
                Si abundance                   &  0.11              &$0.25\pm0.05$  \\
                $N_H$ ($10^{22}$ cm$^{-2}$)    &   0.75                & $ 0.81\pm0.12$  \\
\enddata
\tablenotetext{a}{In 0.7 -- 10 keV.}
\tablecomments{Errors correspond to 90 \% confidence.}
\label{NEI_fitting}
\end{deluxetable}
\end{center}

\clearpage
\begin{center}
\begin{deluxetable}{cccccc}
\tabletypesize{\footnotesize}
\tablecaption{Comparison of the 2--10 keV GRXE fluxes by different satellites}
\tablehead{\colhead{Satellite}&\colhead{GRXE flux$^a$}                      & \colhead{Threshold flux$^a$}                &\colhead{Region}&\colhead{Area}   & \colhead{Reference}\\
           \colhead{}         &\colhead{(erg s$^{-1}$ cm$^{-2}$ deg$^{-2}$)}& \colhead{(erg s$^{-1}$ cm$^{-2}$)}&\colhead{}      &\colhead{(deg$^2$)}&\colhead{}
}
\startdata
{\em ASCA}   &  $5.5 \times 10^{-11}$$^{b,c}$                                       & $5  \times 10^{-13}$$^d$          &$l\approx28.5,-0.5\lesssim b \lesssim2$  &  3.3         &Kaneda et al.\ (1997)\\
{\em ASCA}   &  $5.2 \times 10^{-11}$$^b$                         & $ 10^{-11}$                       &$|l|\lesssim35,|b|\lesssim0.4$       &    63           &Sugizaki et al.\ (2001)\\
{\em XMM}    &  $9.6 \times 10^{-11}$                         & $2  \times 10^{-12}$              &$19\lesssim l\lesssim22,|b|\lesssim0.6$       & 3             &Hands et al.\ (2004)\\
{\em XMM}    &  $9.3 \times 10^{-11}$                         & $2  \times 10^{-13}$              &$19\lesssim l\lesssim22,|b|\lesssim0.6$       & 3             &Hands et al.\ (2004)\\
{\em Chandra}&  $7.4 \times 10^{-11}$                         & $2  \times 10^{-13}$              &$(l,b)\approx(28.5,0.0)$ & 0.07          & This work\\
{\em Chandra}&  $6.5 \times 10^{-11}$                         & $3  \times 10^{-15}$              &$(l,b)\approx(28.5,0.0)$ & 0.07          & This work\\
\enddata
\tablenotetext{a}{Sources brighter than the threshold flux are excluded in the calculation of the GRXE flux
which includes point sources below the threshold  and the diffuse emission.}
\tablenotetext{b}{Assuming a uniform spatial distribution with $1.\arcdeg5$  radius.}
\tablenotetext{c}{Calculated from the best-fit spectral parameters in the paper.}
\tablenotetext{d}{Taken from Yamauchi et al. (1996) in the same observation area.}

\label{compare_xmm_asca}
\end{deluxetable}
\end{center}
\end{document}